\documentclass[aps,prd,reprint,twocolumn,10pt,eqsecnum,showpacs,nofootinbib,longbibliography,superscriptaddress,floatfix]{revtex4-2}

\usepackage{amsmath}
\usepackage{amssymb}
\usepackage{amsfonts}
\usepackage{mathrsfs}
\usepackage{mathtools}
\usepackage{multirow}

\usepackage{hyperref}

\usepackage[section]{placeins}
\usepackage{accents}

\usepackage{color}

\newcommand{\ud}{\mathrm{d}}

\newcommand{\scri}{\mathscr{I}}
\newcommand{\scrh}{\mathscr{H}}
\newcommand{\lie}{\pounds}
\newcommand{\var}{\delta}

\newcommand{\interchange}[2]{#1 \longleftrightarrow #2}
\newcommand{\bs}[1]{\boldsymbol{#1}}
\newcommand{\mc}[1]{\mathcal{#1}}
\newcommand{\ms}[1]{\mathscr{#1}}
\newcommand{\ul}[1]{\underline{#1}}
\newcommand{\<}{\langle}
\renewcommand{\>}{\rangle}

\allowdisplaybreaks[3]
\begin{document}

\title{Flux-balance laws for spinning bodies under the gravitational self-force}
\author{Alexander M. Grant}
\email{a.m.grant@soton.ac.uk}
\affiliation{School of Mathematical Sciences and STAG Research Centre, University of Southampton,
  Southampton, SO17 1BJ, United Kingdom}

\begin{abstract}
  The motion of an extended, but still weakly gravitating body in general relativity can often be determined by a set of conserved quantities.
  Much like for geodesic motion, a sufficient number of conserved quantities allows the motion to be solved by quadrature.
  Under the gravitational self-force (relaxing the ``weakly gravitating'' assumption), the motion can then be described in terms of the evolution these ``conserved quantities''.
  This evolution can be calculated using the (local) self-force on the body, but such an approach is computationally intensive.
  To avoid this, one often uses \emph{flux-balance laws}: relationships between the average evolution (capturing the dissipative dynamics) and the values of the field far away from the body, which are far easier to compute.
  In the absence of spin, such a flux-balance law has been proven in~\cite{Isoyama2018} for any of the conserved action variables appearing in a Hamiltonian formulation of geodesic motion in the Kerr spacetime.
  In this paper, we derive a corresponding flux-balance law, to linear order in spin, directly relating average rates of change to the flux of a conserved current through the horizon and out to infinity.
  In the absence of spin, this reproduces results consistent with those in~\cite{Isoyama2018}.
  To linear order in spin, we construct flux-balance laws for four of the five constants of motion for spinning bodies in the Kerr spacetime, although not in a practical form.
  However, this result provides a promising path towards deriving the flux-balance law for the (generalized) Carter constant.
\end{abstract}

\maketitle

\tableofcontents

\section{Introduction} \label{sec:intro}

Only idealized point particles in general relativity obey the equivalence principle and follow geodesics in the background spacetime.
More general bodies will fail to follow geodesics for two reasons: either they are sufficiently extended that they cannot be characterized only by their mass and four-velocity, requiring spin and higher multipole moments, or they are not sufficiently weakly gravitating, so that they cannot be treated as test bodies in some fixed background spacetime.
When the test body approximation is valid, the motion is described by the Mathisson-Papapetrou-Dixon equations~\cite{Mathisson1937, Papapetrou1951, Dixon1970a, Dixon1970b, Dixon1974}.
By going beyond the test-body limit, one needs to include the effects of the gravitational \emph{self-force}.

While the self-force is often neglected, there exist astrophysical regimes where the gravitational self-force is quite relevant.
One such regime is in an extreme mass-ratio inspiral (EMRI): a stellar-mass compact object (which we call the secondary, of mass $m$) orbiting a supermassive black hole (the primary, of mass $M \gg m$).
The mass ratio, $\varepsilon \equiv m/M$, measures how strongly the secondary affects the surrounding spacetime, and is therefore a measure of the strength of the gravitational self-force.
Due to their low-frequency nature, gravitational waves emitted by these systems are expected to be an important source for the upcoming space-based detector LISA~\cite{Amaro-Seoane2017, Babak2017, Gair2017, LISA2022}.
However, in addition to the ever-present noise, there will be many different sources that all contribute to the signal seen by LISA~\cite{Amaro-Seoane2017, LISA2022}.
The computation of detailed, efficiently-produced gravitational wave models for EMRI signals will be important for their detection, parameter estimation, and any tests of general relativity that they may provide.

It is with this goal in mind that, in recent years, there has been much progress in understanding the gravitational self-force; see for example~\cite{Barack2018, Pound2021} for reviews of the more pragmatic aspects and progress on the computation of waveforms, or more foundational reviews in~\cite{Poisson2011, Harte2014, Pound2015b}.
While much of the work has focused on the simplest case, where the secondary has vanishing spin and higher multipoles, astrophysically relevant EMRIs may have secondaries with non-negligible spin.
Failure to capture these effects could result in biases in parameter estimation, or in spurious ``violations'' of general relativity, and so there have been significant recent efforts to compute the effects of spin on self-forced motion (see, for example,~\cite{Mathews2021, Drummond2022a, Drummond2022b, Drummond2023, Skoupy2021, Skoupy2022, Skoupy2023, Skoupy2024}).

For black holes, the spin is constrained to be smaller than the square of the mass, and so the effects of spin are second-order in the mass ratio, and for bodies with $O(1)$ compactness, this will also be the case (this is why the spin appears at second order in the discussion of~\cite{Pound2015b}).
Because one would also need to generically include the effect of higher spin-induced multipoles (as in Newtonian theory; see for example Sec.~2.4.5 of~\cite{Poisson2014}), we truncate at linear order in spin, and neglect all higher multipoles, which are similarly highly suppressed for compact bodies.
Despite the fact that it is somewhat inconsistent, in this paper we neglect the full effects of second-order self-force, leave such a discussion to future work, and consider the effects of spin to be of part of the first-order self-force.

Up to certain caveats laid out in Footnote~\ref{foot:caveat} below, the motion in this case is described by a sort of generalized equivalence principle: the motion of a self-gravitating extended body is the same as that of a test extended body in an ``effective'' metric $\breve g_{ab}$.
This effective metric is constructed from the background metric $g_{ab}$ and a regularized metric perturbation at the location of the body.
While this is certainly a simple prescription, \emph{solving} for the motion in this effective metric is computationally intensive, due to difficulties in constructing this regularized metric perturbation, or indeed any metric perturbation, at the location of the body (see, for example, the extensive discussion in~\cite{Pound2021}).

A more tractable approach, using \emph{flux-balance laws}, has often been considered throughout the literature.
Here, instead of solving for this effective metric near the body to get the \emph{exact} motion of the body, a flux-balance law relates the \emph{average} motion of the body to the emitted radiation that reaches the boundaries of the system: infinity and the horizon of the black hole.
An example of a flux-balance law appearing in physics folklore is in the decay of the classical atom: since accelerating charges radiate, the emission of energy by an orbiting electron should decrease its orbital energy, resulting in a plunge towards the nucleus.
This heuristic argument can also be developed in general relativity in the widely-separated, slow-motion limit, with varying levels of rigor (see~\cite{Damour1983, Blanchet1993} and references therein, and~\cite{Blanchet2024} for more recent developments up to 4.5 post-Newtonian order).
Such flux-balance laws have provided a crucial test of general relativity from the measurement of the changing orbital period of the Hulse-Taylor binary pulsar~\cite{Taylor1982}.

These examples highlight a key feature of flux-balance laws: they allow for the direct computation not of (even averaged) position or velocity, but of the slowly-evolving, no-longer-conserved quantities of the background motion.
In the Kerr spacetime, for example, there are four constants of motion for a geodesic: the square of the mass $m^2$, the energy $E$, the azimuthal angular momentum $L_z$, and the ``Carter constant'' $K$~\cite{Carter1968a}.
The Carter constant, which like $m^2$ is quadratic in the four-momentum, is related to the square of the orbital angular momentum $L^2$ in the limit where the primary black hole has no spin.
These are sufficient to determine geodesic motion, but when the body has spin, in addition to generalizations of each of these constants~\cite{Rudiger1981, Rudiger1983}, there exists a fifth constant of motion, which we call the R\"udiger constant.\footnote{Elsewhere in the literature (for example,~\cite{Compere2021b, Ramond2022b}), this constant is referred to as the \emph{linear} R\"udiger invariant, in contrast to the generalization of the Carter constant mentioned above, which is often called the \emph{quadratic} R\"udiger invariant.
  For simplicity, we instead use the term ``Carter constant'' to refer to both the usual constant of motion for non-spinning particle motion and its generalization to spinning-body motion.
  As such, ``R\"udiger constant'' in this paper refers only to the constant of motion which is linear in the four-momentum and spin of the body.} $Y$~\cite{Rudiger1981}
For any sort of flux-balance law to be useful, it must therefore provide average rates of change of these quantities.

Two general approaches have been used in the literature in order to derive flux-balance laws.
The first is to use the properties of Green's functions in the Kerr spacetime to derive relationships between the rates of change of quasi-conserved quantities to the amplitudes of solutions to the Teukolsky equation, which can be written as terms defined at the horizon and infinity.
This approach was pioneered in~\cite{Galtsov1982} for computations of $E$ and $L_z$ without spin, which was then (at a much later date) extended to the linear-in-spin case in~\cite{Akcay2019} (whose derivation was later corrected in~\cite{Mathews2021}).
In order to determine the full motion without spin, the evolution of the Carter constant $K$ was computed (once again in terms of asymptotic quantities) in~\cite{Sago2005a, Sago2005b}, which was later extended in~\cite{Isoyama2018} to the case of \emph{arbitrary} constants of motion for geodesic motion in Kerr.
The conserved quantities considered in~\cite{Isoyama2018} were the action variables in a Hamiltonian formulation of geodesic motion, which they then show can be used to compute the evolution of $K$.

The second approach to flux-balance laws is more physically motivated: much like in the ``classical atom'' case above, the asymptotic quantities are the fluxes of a conserved current (there, the Poynting vector) through the horizon and null infinity.
While such an idea is hinted at in~\cite{Galtsov1982}, such a flux-balance law was only rigorously derived in~\cite{Quinn1999} (without spin), for the case of conserved quantities defined from spacetime symmetries, such as $E$ and $L_z$.
However, this left the physical intuition for the flux-balance law for $K$ from~\cite{Sago2005a, Sago2005b} opaque: the conserved currents considered in~\cite{Quinn1999} involved the stress-energy tensor, and there was no known way to derive a conserved current from the stress-energy tensor and the Killing tensor $K_{ab}$ from which $K$ was defined~\cite{Walker1970}.
Moreover, it was even shown that no conserved current could be constructed from $K_{ab}$ and $T^{ab}$ which had all of the desired properties~\cite{Grant2015}.

Recently, however, it was shown in~\cite{Grant2022b} that, at least in the context of a scalar field theory and in the absence of spin (where the evolution of the Carter constant was derived in~\cite{Drasco2005}), one can write the average evolution of the action variables in terms of fluxes of an appropriate conserved current through a local worldtube around the body.
This presented a partial physical explanation for the results of~\cite{Isoyama2018}.
Moreover, by na\"ively guessing how to extend this flux-balance law all the way to the horizon and null infinity, a derivation was presented in~\cite{Grant2022b} for a flux-balance law, for scalar fields, that was exactly analogous to the results in~\cite{Isoyama2018}.

This paper begins where~\cite{Grant2022b} ends, and provides an extension of this local flux-balance law beyond the limited scope of that paper.
As such, the goals of this paper are threefold:
\begin{itemize}

\item to show that a local flux-balance law also holds for gravity;

\item to extend this into a global flux-balance law, defined in terms of fluxes through the horizon and null infinity; and

\item to show that the global flux-balance law \emph{also} exists, at least in a limited sense, for a spinning body to linear order in spin.

\end{itemize}

The main result of this paper is the following formula, which holds generically for any black hole spacetime (that is, a spacetime possessing a horizon $\scrh$ and null infinity $\scri$), for a spinning body (to linear order in spin) undergoing first-order self-forced motion:\footnote{In general, there are situations where neither the left- nor right-hand sides of this equation are well defined, which we will discuss in more detail in Secs.~\ref{sec:flux_balance} and~\ref{sec:geodesic_kerr} below.}
\begin{equation} \label{eqn:result}
  \left\<\var \dot \Gamma^A (\tau')\right\>_{\tau'} = \Pi^{AB} \left[\mc F_B + \nabla_B \<\mc H(X')\>_{\tau'}\right],
\end{equation}
Without describing the notation explicitly, we will briefly describe this equation's general features: first, the left-hand side represents a notion of an average rate of change of quantities defined on the worldline, and is defined in Eqs.~\eqref{eqn:avg_def} and~\eqref{eqn:avg_vardot}.
For the right-hand side, the first term is a flux of a conserved current through null infinity and the horizon that is defined in terms of metric perturbations in Eq.~\eqref{eqn:boundary_fluxes} below, and the second term is a derivative of an average ``conservative'' Hamiltonian (defined in Sec.~\ref{sec:conservative}) which generically appears, and is important in the case of resonances~\cite{Isoyama2018}.

Unfortunately, because of the presence of the conservative Hamiltonian, this equation does \emph{not} constitute a flux-balance law relating rates of change on the worldline to asymptotic fluxes of a conserved current, as this term still involves computing the regularized metric perturbation, from which the effective metric $\breve g_{ab}$ is constructed, on the worldline.
Moreover, the left-hand side is not directly related to averaged rates of change of perturbed conserved quantities.
As described in more detail in Sec.~\ref{sec:formalism} below, it is a covariantly-defined average of the rate of change of $\var \Gamma^A$, the first-order deviation of the self-forced motion from the background trajectory $\Gamma$ through phase space.
However, in order to make this average covariantly defined on phase space [like the remaining terms in Eq.~\eqref{eqn:result}], one needs to include a bitensor introduced in~\cite{Grant2022b} called the \emph{Hamilton propagator}, which significantly complicates the interpretation in the case of spin, as we discuss in Sec.~\ref{sec:spinning_kerr} below.

However, in the specific case where the motion on phase space is integrable, namely there exists a sufficient number of independent constants of motion $P_\alpha$ that one can construct action-angle variables on phase space, Eq.~\eqref{eqn:result} simplifies, yielding:
\begin{equation} \label{eqn:true_result}
  \left\<\var \dot P_\alpha (\tau')\right\>_{\tau'} = -A^\beta{}_\alpha (\partial_{\vartheta^\beta})^A \mc F_A,
\end{equation}
assuming that the background motion is non-resonant.
Here $\vartheta^\alpha$ represents the angle variables, $A^\beta{}_\alpha$ describes the failure of the coordinates $(\vartheta^\alpha, P_\alpha)$ to be canonical, and $(\partial_{\vartheta^\beta})^A \mc F_A$ is the angle-variable component of $\mc F_A$.
The free index of $\mc F_A$ arises from a derivative acting on the metric perturbation, and so this equation implies that it is the derivative of the metric perturbation with respect to the angle variables $\vartheta^\alpha$ (in a manner which we make more precise through the body of this paper) which is relevant to computing the evolution of the constants of motion $P_\alpha$.

Unlike Eq.~\eqref{eqn:result}, Eq.~\eqref{eqn:true_result} is a \emph{true} flux-balance law, as the right-hand side only involves the computation of the metric perturbation $h_{ab}$ at null infinity and the horizon.
Moreover, the left-hand side is the average rate of change of the perturbed conserved quantity $\var P_\alpha$, in the usual sense.
In particular, this equation holds for non-spinning motion in the Kerr spacetime, and is equivalent to the results of~\cite{Isoyama2018}.
However, in the case of spin, there no longer exists a full set of action-angle variables on the entire phase space, in a sense which we describe in more detail in Sec.~\ref{sec:spinning_kerr} below.
As such, our results are much more limited: while we can still recover flux-balance laws for generalizations of $m^2$, $E$, $L_z$, and $K$, the evolution of the R\"udiger constant $Y$ is less well-behaved.
As mentioned above, while flux-balance laws for $E$ and $L_z$ have already been known~\cite{Akcay2019, Mathews2021}, the flux-balance law for $K$ is new to this work, and arises due to a ``miracle'' that we describe in more detail in Sec.~\ref{sec:spinning_kerr} below.
While the results of this paper are somewhat abstract, they provide a path towards explicit expressions for these flux-balance laws, as discussed in~\cite{Mathews2025} using recent results in~\cite{Witzany2024}.

The structure of the remainder of this paper is as follows.
First, in Sec.~\ref{sec:hamiltonians} we provide a review of pseudo-Hamiltonian systems, a generalization of Hamiltonian systems that allows for the presence of dissipative effects, such as the gravitational self-force.
As can be seen above, all of the results of this paper can be naturally described in terms of quantities on phase space, and so this is the natural arena in which to tackle the problem of flux-balance laws.
Next, in Sec.~\ref{sec:fields}, we discuss three aspects of perturbative field theory that are necessary for later derivations: a conserved current (the symplectic current) which we use to construct our fluxes at the boundaries of the system, the field equations for the metric perturbations that appear in the gravitational self-force, and a relationship between the pseudo-Hamiltonian in Sec.~\ref{sec:hamiltonians} and the stress-energy tensor sourcing these metric perturbations which we call the \emph{Hamiltonian alignment condition}.
Putting together the results of Secs.~\ref{sec:hamiltonians} and~\ref{sec:fields}, we derive a precursor to Eq.~\eqref{eqn:result} in Sec.~\ref{sec:flux_balance}, for arbitrary spacetimes.
We then consider applications to the Kerr spacetime in Sec.~\ref{sec:kerr}, and both derive Eq.~\eqref{eqn:true_result} for the motion of a non-spinning body and discuss difficulties in generalizing to linear order in spin.
We conclude with a summary and discussion of future work in Sec.~\ref{sec:discussion}.

We use the following conventions and notation in this paper.
First, we use the mostly plus signature for the metric, and we use the conventions for the Riemann tensor and differential forms of Wald~\cite{Wald1984}.
Moreover, we follow the convention of using lowercase Latin letters from the beginning of the alphabet ($a$, $b$, etc.) to denote abstract indices on the spacetime manifold $M$, with Greek letters ($\alpha$, $\beta$, etc.) denoting spacetime coordinate indices, and hatted Greek letters ($\hat \alpha$, $\hat \beta$, etc.) for component indices along an orthonormal tetrad.
Following~\cite{Penrose1987}, we use script capital Latin letters ($\ms A$, $\ms B$, etc.) for general ``composite'' indices denoting a collection of abstract indices.

Following~\cite{Grant2022b}, for tensor fields on the phase space manifold $\mc M$, we use capital Latin letters from the beginning of the alphabet ($A$, $B$, etc.) to denote abstract indices, and capital Hebrew letters from the beginning of the alphabet ($\aleph$, $\beth$, etc.) for coordinate indices.
We will also use unhatted Greek letters for indices associated with groupings of four coordinates on phase space, such as the angle variables $\vartheta^\alpha$ or the conserved quantities $P_\alpha$.
A difficulty arises then in denoting the components of tensors along these directions: as in Eq.~\eqref{eqn:true_result} above, we resort to contracting tensors with (for example) the vector fields $(\partial_{\vartheta^\alpha})^A$ and $(\partial_{P_\alpha})^A$, which form a basis dual to the usual one-forms $(\ud \vartheta^\alpha)_A$ and $(\ud P_\alpha)_A$.
Finally, for quantities appearing on the spacetime manifold $M$, we typically denote corresponding quantities on the phase space manifold $\mc M$ with capitalized versions of the same symbol; for example, the worldline $\gamma$ is determined by a trajectory $\Gamma$ through phase space.

We use the notation for bitensors from~\cite{Poisson2011}, and we use a convention where indices at some point will have the same adornments as the point itself, and we do not explicitly denote the dependence of a bitensor on the point unless it is a scalar at that point (in particular, if the indices being used are coordinate indices).
As such, for example, $\omega^{a'}{}_{A''} (x)$ denotes a vector at $x' \in M$, a one-form at $X'' \in \mc M$, and a scalar at $x \in M$; its components would be denoted by $\omega^\alpha{}_\aleph (x', X'', x)$.
We will also occasionally use an index-free notation, either for (tensor-valued) differential forms on the spacetime manifold or for tensorial arguments to functionals, and describe where the ``invisible'' indices lie if it is ambiguous.
Tensors in this index-free notation will be denoted in bold to make it clear that indices are being suppressed.

Finally, parentheses and square brackets will typically have their conventional meanings (that is, dependence as a function vs. dependence as a functional/coincidence limits~\cite{Poisson2011}), to which we add the additional notation that curly braces will be used to indicate dependence as a local, linear functionals (from~\cite{Grant2022a}).
In certain situations, however, when brackets of the same type would be nested, such as in $f(g(x))$, we follow the usual convention of cycling through alternative brackets; in this example, we would use $f[g(x)]$, even if $f$ is not a functional of its argument.
That this replacement of brackets has occurred should be clear from the surrounding context.

\section{(Pseudo-)Hamiltonian formulations} \label{sec:hamiltonians}

In this section, we first review Hamiltonian and pseudo-Hamiltonian formulations, which can be used to describe the background and perturbed motion under the gravitational self-force, respectively.
We then describe how this is explicitly done in the case of motion to linear order in spin.

\subsection{General framework} \label{sec:formalism}

We start by briefly reviewing Hamiltonian systems.
Here, one starts with a phase space manifold $\mc M$, which has the structure of a fiber bundle over the spacetime manifold $M$.
For example, for point particle motion, $\mc M$ is the cotangent bundle.
Because it is a fiber bundle, there is a natural projection map $\pi: \mc M \to M$.
The worldline of the body, which is given by a curve $\gamma$ through $M$, is determined by the projection of a trajectory $\Gamma$ through $\mc M$:
\begin{equation}
  \gamma = \pi \circ \Gamma.
\end{equation}
We denote the common parameter of $\gamma$ and $\Gamma$ by $\tau$, as it will represent the proper time of the body, so that $\dot \gamma^a$ will be normalized.

We further assume that $\mc M$ is a \emph{Poisson manifold} (for a review, see Chapter 10 of~\cite{Marsden1999}): there exists an antisymmetric \emph{Poisson bivector} $\Pi^{AB}$, defining a Poisson bracket $\{\cdot, \cdot\}$ for scalar fields $f$ and $g$ by
\begin{equation} \label{eqn:poisson}
  \{f, g\} \equiv \Pi^{AB} (\ud f)_A (\ud g)_B,
\end{equation}
that satisfies the Jacobi identity:
\begin{equation}
  \{f, \{g, h\}\} + \{g, \{h, f\}\} + \{h, \{f, g\}\} = 0;
\end{equation}
this is equivalent to the statement that the Schouten(-Nijenhuis) bracket of $\Pi^{AB}$ with itself vanishes (see Theorem~10.6.2 of~\cite{Marsden1999}).
The equations of motion for $\gamma$ are then given by a Hamiltonian formulation if there exists a function $H$ on $\mc M$ that satisfies Hamilton's equations:
\begin{equation} \label{eqn:hamilton}
  \dot \Gamma^A = \Pi^{AB} \nabla_B H.
\end{equation}
Here, $\nabla_A$ is any covariant derivative on phase space; we use this notation, instead of the exterior derivative notation $(\ud H)_A$, as it is clearer when we discuss pseudo-Hamiltonians below.
Note that we do not assume that $\mc M$ is a \emph{symplectic} manifold, which means that $\Pi^{AB}$ can potentially be degenerate.
As we will describe below, while $\Pi^{AB}$ is nondegenerate for point particle motion, it is degenerate at linear order in spin.

Now, we consider the more general case of a \emph{pseudo}-Hamiltonian system.
Here, the idea is that what one would typically think of as a Hamiltonian is not simply a function of a point on phase space $X$, but a function also of a trajectory through phase space $\bar \Gamma$ which is determined by a point $\bar X$ through which $\bar \Gamma$ passes.
Following~\cite{Blanco2022, Blanco2023, Blanco2024}, we call such a function a \emph{pseudo-Hamiltonian}.
In the pseudo-Hamiltonian generalization of Hamilton's equations, one takes a derivative with respect to $X$ at $\bar X$ fixed, and then takes $\bar X$ and $X$ to lie along the same trajectory $\Gamma$ in phase space.
These systems allow for dissipation, unlike ``true'' Hamiltonian systems~\cite{Galley2012}.\footnote{Note that, while they sound somewhat exotic, the idea of pseudo-Hamiltonian systems is far from new (see the discussion and references in~\cite{Galley2012}).
  For an example from the gravitational radiation literature, the Burke-Thorne potential for the leading-order post-Newtonian radiation-reaction force (see the discussion in Chapter~12 of~\cite{Poisson2014}),
  \[
    \Phi^{RR} (x) = \frac{1}{5} x_i x_j \frac{\ud^5 I^{\<ij\>}}{\ud t^5},
  \]
  can be thought of as the potential energy part of a pseudo-Hamiltonian: the quadrupole moment $I^{ij}$ depends on the trajectory of the object undergoing radiation reaction, which is kept fixed when $\Phi^{RR}$ is differentiated with respect to $x^i$ to find the force \{to see this, compare Eqs.~(12.198) and~(12.204) of~\cite{Poisson2014}\}.}

Explicitly, consider the Hamiltonian flow map $\Upsilon_\Delta: \mc M \to \mc M$ associated with the background motion.
This map is defined as follows: given some $X \in \mc M$, suppose that
\begin{equation}
  X = \Gamma(\tau),
\end{equation}
for some value of $\tau$ along some trajectory $\Gamma$ in $\mc M$ that passes through $X$ and obeys Hamilton's equations for the background Hamiltonian $H$.
We then define
\begin{equation}
  \Upsilon_\Delta (X) = \Gamma(\tau + \Delta).
\end{equation}
The pseudo-Hamiltonian $H(\varepsilon)$ is a function of two points, $X$ and $\bar X$, such that
\begin{equation}
  H(X, \bar X; \varepsilon) = H[X, \Upsilon_\Delta (\bar X); \varepsilon],
\end{equation}
where $\Delta \in \mathbb R$ is arbitrary.
This captures the idea that the dependence on $\bar X$ is only through the trajectory determined by it.
Moreover, we require that
\begin{equation}
  H(X, \bar X; 0) = H(X);
\end{equation}
that is, when the small parameter $\varepsilon = 0$, the system becomes Hamiltonian for the background motion that determines $\Upsilon$.\footnote{In order to describe self-force in the so-called ``self-consistent formalism''~\cite{Pound2009, Pound2015b}, one would need to instead consider the Hamiltonian flow map associated with the exact motion in the definition of the pseudo-Hamiltonian, which we would denote $\Upsilon_\Delta (\varepsilon)$.
  However, since we are computing the first-order self-force, the self-consistent formalism is not necessary, and so we just use the background Hamiltonian flow map.}

Note that $X$ and $\bar X$ are completely disconnected as variables on which our pseudo-Hamiltonian depends; it is only when we determine the trajectory that we necessarily demand (for self-consistency) that $\bar X$ be along the curve that passes through $X$.
As such, for an arbitrary parameter $\Delta$, we define the trajectory to be such that the tangent vector satisfies
\begin{equation} \label{eqn:pseudohamilton}
  \dot \Gamma^A (\varepsilon) = \Pi^{AB} (\varepsilon) [\nabla_B H(X, \bar X; \varepsilon)]_{\bar X \to \Upsilon_\Delta (X)}.
\end{equation}
The expression on the right-hand side is a \emph{coincidence limit} of the bitensorial expression in brackets, which is defined simply by taking the limit $\bar X \to \Upsilon_\Delta (X)$ after taking the derivative.
Coincidence limits arise naturally in the discussion of bitensors; see, for example, the description in~\cite{Poisson2011}.
Note that this is a slight generalization of the expressions presented in~\cite{Blanco2022, Blanco2023, Blanco2024}, where it was assumed that $\Delta = 0$; we adopt this more general approach as it will simplify the discussion later in Sec.~\ref{sec:flux_balance}.

At this point, we now explicitly restrict to the first-order formulation; here, it is the case that everything which we need to know about the evolution of the perturbed trajectory $\Gamma$ is given by $\var \Gamma^A$, the tangent to the congruence $\Gamma(\tau; \varepsilon)$ where $\tau$ is fixed and $\varepsilon$ varies.
That is, for any scalar field $f$,
\begin{equation}
  \left.\var \Gamma^A \nabla_A f\right|_{\Gamma(\tau)} \equiv \left.\frac{\partial f[\Gamma(\tau; \varepsilon)]}{\partial \varepsilon}\right|_{\varepsilon = 0}.
\end{equation}
As we describe more explicitly in Sec.~\ref{sec:fields} below, let us also define the variation of any tensor $\bs Q$ by
\begin{equation} \label{eqn:var}
  \var \bs Q \equiv \left.\frac{\partial \bs Q}{\partial \varepsilon}\right|_{\varepsilon = 0}.
\end{equation}
As shown in Sec.~II.B.2 of~\cite{Grant2022b}, the following evolution equation for $\var \Gamma^A$ follows from Eq.~\eqref{eqn:pseudohamilton}, using the notation of the pseudo-Hamiltonian formalism (which was only implicit in~\cite{Grant2022b}):
\begin{equation} \label{eqn:hamilton_pert}
  \lie_{\dot \Gamma} \var \Gamma^A = \Pi^{AB} [\nabla_B \var H(X, \bar X)]_{\bar X \to \Upsilon_\Delta (X)},
\end{equation}
where we have also explicitly assumed that (as in~\cite{Grant2022b})
\begin{equation}
  \var \Pi^{AB} = 0.
\end{equation}
That is, the Poisson bracket structure is \emph{fixed}\footnote{When we say ``fixed'' here, we mean that it does not \emph{explicitly} depend on $\varepsilon$; any possible implicit dependence due to being evaluated at a point which itself depends on $\varepsilon$ is allowed, as can be seen in the derivation in~\cite{Grant2022b}.} as a function of $\varepsilon$.
Note that this is a ``gauge-dependent'' statement: it only holds in a certain class of coordinates [for example, those in Eq.~\eqref{eqn:good_coords}], and if one performs an $\varepsilon$-dependent coordinate transformation (depending, for example, on the metric perturbation), then Eq.~\eqref{eqn:hamilton_pert} will necessarily change.
It is for this reason that we work in coordinates related to those in Eq.~\eqref{eqn:good_coords} by field-independent coordinate transformations.

In order to solve Eq.~\eqref{eqn:hamilton_pert}, one needs to integrate vectors on phase space.
On a general manifold, one does not have a means of adding and subtracting vectors at different points, which is necessary in order to have a well-defined notion of integration along a curve.
On the spacetime manifold, one can typically use parallel transport, that is, the parallel propagator bitensor $g^{a'}{}_a$, but there is no natural notion of parallel transport on phase space.
However, on phase space there still exists a bitensor that allows for transport for vectors along $\Gamma$, the \emph{Hamilton propagator}~\cite{Grant2022b}.
This is a bitensor defined along $\Gamma$ by the following equation: for any scalar field $f$, and any pair of points $X = \Gamma(\tau)$ and $X' = \Gamma(\tau')$, the Hamilton propagator $\Upsilon^{A'}{}_A$ satisfies
\begin{equation} \label{eqn:pushforward}
  \nabla_A (f \circ \Upsilon_{\tau' - \tau}) = \Upsilon^{A'}{}_A \nabla_{A'} f.
\end{equation}
In other words, the Hamilton propagator is the \emph{pushforward} of the Hamiltonian flow map $\Upsilon_\Delta$,\footnote{While the pushforward of a mapping from a manifold into itself is not discussed in terms of bitensors in typical textbooks on general relativity (such as, for example, Appendix~C of~\cite{Wald1984}), it is not unprecedented: see for example Sec.~1.4 of~\cite{Marsden1983}.} and is a bitensor that is a one-form at $\Gamma(\tau)$ and a vector at $\Gamma(\tau + \Delta)$.\footnote{Note that this definition is slightly different from that originally presented in~\cite{Grant2022b}: there, $\Upsilon^{A'}{}_A$ was defined such that it was a bitensor at \emph{any} two points $X$ and $X'$ along a given trajectory $\Gamma$, by using a function which gave the difference $\Delta$ in proper times between these two points.
  This allows one to take derivatives of $\Upsilon^{A'}{}_A$ with respect to either of $X$ and $X'$, entirely independently.
  However, as we discovered while writing the present paper, such a feature is not required for the derivation of flux-balance laws, and so we do not include this part of the definition.}

In coordinates, the chain rule implies that
\begin{equation} \label{eqn:coordinates}
  \Upsilon^\aleph{}_\beth (\tau', \tau) = \frac{\partial X^\aleph (\tau')}{\partial X^\beth (\tau)},
\end{equation}
where $X^\aleph (\tau)$ denotes the coordinates of $\Gamma(\tau)$.
That is, in order to compute this bitensor, and so integrate Eq.~\eqref{eqn:hamilton_pert}, one needs to solve for the background trajectory in terms of initial data.
We perform this explicit calculation in the relevant cases in Secs.~\ref{sec:geodesic_kerr} and~\ref{sec:spinning_kerr}.

In addition to Eqs.~\eqref{eqn:pushforward} and~\eqref{eqn:coordinates}, the Hamilton propagator has a few very convenient properties.
First, as it is created from a map from the manifold into itself, it has a natural composition property:
\begin{equation} \label{eqn:composition}
  \Upsilon^A{}_{A'} \Upsilon^{A'}{}_{A''} = \Upsilon^A{}_{A''}, \quad \Upsilon^A{}_{A'} \Upsilon^{A'}{}_B = \delta^A{}_B.
\end{equation}
Next, consider the Lie derivative with respect to $\dot \Gamma^A$ of some tensor $T^{\ms A}$, where $\ms A$ is some composite index.
For such composite indices, there exists a Hamilton propagator $\Upsilon^{\ms A'}{}_{\ms A}$ which uses $\Upsilon^{A'}{}_A$ for mapping contravariant indices at $X$ to $X'$, and $\Upsilon^A{}_{A'}$ for covariant indices.
The Lie derivative is then defined by \{compare to Eq.~(C.2.1) of~\cite{Wald1984}\}
\begin{equation}
  \lie_{\dot \Gamma} T^{\ms A} \equiv \lim_{\lambda \to 0} \left.\frac{\ud}{\ud \lambda} \left(\Upsilon^{\ms A}{}_{\ms A'} T^{\ms A'}\right)\right|_{X' = \Gamma(\tau + \lambda)}.
\end{equation}
As such, from Eq.~\eqref{eqn:composition} it follows that
\begin{equation} \label{eqn:lie_ode}
  \begin{split}
    \Upsilon^{\ms A}{}_{\ms A'} \lie_{\dot \Gamma'} T^{\ms A'} &= \lim_{\lambda \to 0} \left.\frac{\ud}{\ud \lambda} \left(\Upsilon^{\ms A}{}_{\ms A''} T^{\ms A''}\right)\right|_{X'' = \Gamma(\tau' + \lambda)} \\
    &= \left.\frac{\ud}{\ud \tau'} \left(\Upsilon^{\ms A}{}_{\ms A'} T^{\ms A'}\right)\right|_{X' = \Gamma(\tau')},
  \end{split}
\end{equation}
by the chain rule.
As such, if the Lie derivative of $T^{\ms A}$ is known at all points along $\Gamma$, one can potentially integrate this equation to solve for $T^{\ms A}$.

We now use the Hamilton propagator to define a notion of an average rate of change of $\var \Gamma^A$: first, define
\begin{equation}
  \var \dot \Gamma^A (\tau') \equiv \Upsilon^A{}_{A'} \lie_{\dot \Gamma'} \var \Gamma^{A'},
\end{equation}
where $X' = \Gamma(\tau')$, and then define, for any bitensor $f^A (\tau')$ that is a scalar at $\tau'$, the average
\begin{equation} \label{eqn:avg_def}
  \left\<f^A (\tau')\right\>_{\tau'} \equiv \lim_{\Delta \tau \to \infty} \frac{1}{\Delta \tau} \int_{\tau_-}^{\tau_+} \ud \tau'\; f^A (\tau'),
\end{equation}
where $\tau_\pm \equiv \tau \pm \Delta \tau/2$.
From this, we have that the average rate of change of $\var \Gamma^A$ is given by
\begin{equation} \label{eqn:avg_vardot}
  \left\<\var \dot \Gamma^A (\tau')\right\>_{\tau'} = \Pi^{AB} \left\<[\nabla_B \var H\{\Upsilon_{\tau' - \tau} (X), \bar X\}]_{\bar X \to X}\right\>_{\tau'},
\end{equation}
where we have used Eq.~\eqref{eqn:pushforward}, together with Eq.~\eqref{eqn:lie_ode} and the fact that
\begin{equation} \label{eqn:lie_pi}
  \lie_{\dot \Gamma} \Pi^{AB} = 0,
\end{equation}
which follows from the Jacobi identity (see Proposition~10.4.1 of~\cite{Marsden1999}).
In principle, one can then use Eq.~\eqref{eqn:lie_ode} to show that
\begin{equation} \label{eqn:avg_diff}
  \left\<\var \dot \Gamma^A (\tau')\right\>_{\tau'} \equiv \lim_{\Delta \tau \to \infty} \frac{\Upsilon^A{}_{A'} \var \Gamma^{A'} - \Upsilon^A{}_{A''} \var \Gamma^{A''}}{\Delta \tau},
\end{equation}
where the points $X'$ and $X''$ in the numerator on the right-hand side are at $\tau \pm \Delta \tau/2$, respectively.
This equation provides an alternative way of thinking about this average rate of change, although we will not use it in this paper.

Finally, we note that, in both Eqs.~\eqref{eqn:avg_vardot} and~\eqref{eqn:avg_diff}, care must be taken that the quantity being averaged does, in fact, have a finite and well-defined average.
As we will see below in Sec.~\ref{sec:geodesic_kerr}, this is not always the case, for particular components of these equations; however, in such cases, it will be true that both the left- and right-hand sides of Eq.~\eqref{eqn:avg_vardot} will be ill-defined.

\subsection{Linear-in-spin motion} \label{sec:pole_dipole}

In this section, we review the motion of spinning test bodies, to linear order in the spin.
The background motion is described in terms of three quantities: the four-velocity $\dot \gamma^a$ (which, for concreteness, we assume is normalized using the metric $g_{ab}$ to have length $-1$), the momentum $p_a$, and the spin $S^{ab}$.
These three quantities obey the Mathisson-Papapetrou-Dixon (MPD) equations~\cite{Mathisson1937, Papapetrou1951, Dixon1970a}:
\begin{subequations} \label{eqn:mpd}
  \begin{align}
    \dot \gamma^b \nabla_b p_a &= -\frac{1}{2} R_{abcd} \dot \gamma^b S^{cd} + F_a, \\
    \dot \gamma^c \nabla_c S^{ab} &= 2 p_c g^{c[a} \dot \gamma^{b]} + N^{ab},
  \end{align}
\end{subequations}
where $F^a$ and $N^{ab}$ are (respectively) forces and torques that are due to the structure of the body, arising from higher multipoles (such as the quadrupole moment, for example).
In this paper, we will neglect the presence of these higher multipoles, applying the \emph{pole-dipole approximation}; due to the generic presence of a spin-induced quadrupole, this requires us to \emph{also} neglect quantities which are second-order in spin, for consistency.

Equations~\eqref{eqn:mpd} do not provide a complete set of equations for the 13 unknowns $\dot \gamma^a$, $p_a$, and $S^{ab}$: these are only 10 equations, and so we must supplement them with an additional choice in order to fix the location of the worldline.
Such a choice is called a \emph{spin-supplementary condition}, and we will adopt the Tulczyjew-Dixon condition~\cite{Tulczyjew1959, Dixon1970a} for the remainder of this paper:\footnote{As we are explicitly excluding the null case, we are free to ignore the known pathologies of this spin-supplementary condition in this case (see, for example, the discussion in~\cite{Deriglazov2017, Harte2022} and pg.~70 of~\cite{Penrose1988}).}
\begin{equation} \label{eqn:tulczyjew}
  d^a \equiv S^{ab} p_b = 0.
\end{equation}
This condition is a constraint that the mass-dipole moment, as measured in the rest frame of $p^a$, vanishes.
Under this spin-supplementary condition, the MPD equations take on the following form, to linear order in spin:
\begin{subequations} \label{eqn:mpd_reduced}
  \begin{align}
    \dot \gamma^a &= \frac{1}{m} g^{ab} p_b + O(\bs S^2), \\
    \dot \gamma^b \nabla_b p_a &= -\frac{1}{2} R_{abcd} \dot \gamma^b S^{cd}, \\
    \dot \gamma^c \nabla_c S^{ab} &= O(\bs S^2),
  \end{align}
\end{subequations}
where
\begin{equation} \label{eqn:m}
  m^2 \equiv -g^{ab} p_a p_b + O(\bs S^2),
\end{equation}
by the normalization of the four-velocity.
For the remainder of this paper, we will implicitly neglect any $O(\bs S^2)$ terms.

Now, one thing to note about the motion described by Eqs.~\eqref{eqn:mpd_reduced} is that it is not, in fact, equivalent to Eqs.~\eqref{eqn:mpd} with the Tulczyjew-Dixon condition satisfied: instead, one can only prove from Eq.~\eqref{eqn:mpd_reduced} that
\begin{equation} \label{eqn:ssc_preserved}
  \dot \gamma^b \nabla_b d^a = 0.
\end{equation}
Instead of considering the ``true'', restricted phase space for spinning bodies in this paper, we will be considering Eq.~\eqref{eqn:mpd_reduced} as given, and then apply $d^a = 0$ as an initial condition, which by Eq.~\eqref{eqn:ssc_preserved} implies that the Tulczyjew-Dixon condition holds for all time.
The Hamiltonian which we consider below is therefore for a ``fictitious'' problem that does not correspond to the physical motion, and the physical motion is only recovered when the initial data are such that Eq.~\eqref{eqn:tulczyjew} holds.
This will be sufficient for our purposes, and (as we will show below in Sec.~\ref{sec:sf_fields}) it is necessary for our results to hold in the form they do.

We now discuss the Hamiltonian formulation of this background motion, starting with the case where the spin vanishes.
This is exactly geodesic motion, given by
\begin{equation}
  \dot \gamma^b \nabla_b \dot \gamma^a = 0.
\end{equation}
As one can straightforwardly show (see, for example,~\cite{Grant2022b}), this equation follows from Hamilton's equations, where the coordinates on the (now 8-dimensional) phase-space are given by
\begin{equation}
  X^\aleph = \begin{pmatrix}
    x^\alpha \\
    p_\alpha
  \end{pmatrix},
\end{equation}
the Poisson bivector $\Pi^{AB}$'s non-zero components are determined by
\begin{equation}
  \{x^\alpha, p_\beta\} = \delta^\alpha{}_\beta,
\end{equation}
and the Hamiltonian is given by
\begin{equation} \label{eqn:hamiltonian_geodesic}
  H(X) = -m(X),
\end{equation}
where we consider the mass $m$, as defined by Eq.~\eqref{eqn:m}, explicitly as a function on phase space.
Hamilton's equations also show that
\begin{equation}
  \dot \gamma^a = \frac{1}{m} g^{ab} p_b,
\end{equation}
directly enforcing the normalization of $\dot \gamma^a$.

For the spinning case, the situation is more subtle.
First, suppose that one uses
\begin{equation}
  X^\aleph = \begin{pmatrix}
    x^\alpha \\
    p_\alpha \\
    S^{\alpha\beta}
  \end{pmatrix}
\end{equation}
as coordinates on the 14-dimensional phase space.
If one uses the Poisson bivector given by
\begin{subequations}
  \begin{align}
    \{x^\alpha, p_\beta\} &= \delta^\alpha{}_\beta, \\
    \{p_\alpha, p_\beta\} &= -\frac{1}{2} R_{\alpha\beta\gamma\delta} S^{\gamma\delta}, \\
    \{S^{\alpha\beta}, p_\gamma\} &= 2 \Gamma^{[\alpha}{}_{\gamma\delta} S^{\beta]\delta}, \\
    \{S^{\alpha\beta}, S^{\gamma\delta}\} &= 2 (g^{\alpha[\delta} S^{\gamma]\beta} - g^{\beta[\delta} S^{\gamma]\alpha})
  \end{align}
\end{subequations}
(a natural choice appearing throughout the literature~\cite{Feldman1980, Tauber1988, dAmbrosi2015}), then the Hamiltonian for the MPD equations with the Tulczjew-Dixon spin supplementary condition is still given by Eq.~\eqref{eqn:hamiltonian_geodesic} (see~\cite{Witzany2018} for an extensive discussion of the relationship between the choice of spin-supplementary condition and Hamiltonian).
This means that $\Pi^{AB}$ is metric-dependent, and as we will see below, this implies that $\var \Pi^{AB} \neq 0$, which significantly complicates the analysis.

Instead, following~\cite{Tauber1988, Khriplovich1989}, we define better coordinates for this problem by first introducing the orthonormal tetrad $\bs e_{\hat \alpha}$, defined by
\begin{equation} \label{eqn:tetrad}
  g_{ab} (e_{\hat \alpha})^a (e_{\hat \beta})^b = \eta_{\hat \alpha \hat \beta}.
\end{equation}
Here, we denote tetrad indices using hatted Greek letters, to distinguish them from spacetime coordinate indices.
The corresponding dual tetrad $\bs e^{\hat \alpha}$ is defined by
\begin{equation}
  (e^{\hat \alpha})_a (e_{\hat \beta})^a = \delta^{\hat \alpha}{}_{\hat \beta}, \qquad (e_{\hat \alpha})^a (e^{\hat \alpha})_b = \delta^a{}_b
\end{equation}
(which are equivalent), from which it follows that
\begin{equation}
  g^{ab} (e^{\hat \alpha})_a (e^{\hat \beta})_b = \eta^{\hat \alpha \hat \beta}.
\end{equation}
Moreover, we consider the connection one-form $\bs \omega_{\hat \alpha \hat \beta}$ defined by
\begin{equation} \label{eqn:spin_connection}
  (\omega_{\hat \alpha \hat \beta})_a = g_{bc} (e_{\hat \alpha})^b \nabla_a (e_{\hat \beta})^c.
\end{equation}
Using these quantities, we define
\begin{equation}
  \hat S^{\hat \alpha \hat \beta} \equiv (e^{\hat \alpha})_a (e^{\hat \beta})_b S^{ab}, \qquad \pi_a \equiv p_a - \frac{1}{2} (\omega_{\hat \alpha \hat \beta})_a \hat S^{\hat \alpha \hat \beta},
\end{equation}
and use the following coordinates on phase space:
\begin{equation} \label{eqn:good_coords}
  X^\aleph = \begin{pmatrix}
    x^\alpha \\
    \pi_\alpha \\
    \hat S^{\hat \alpha \hat \beta}
  \end{pmatrix},
\end{equation}
where $\hat S^{\hat \alpha \hat \beta}$ is antisymmetric.

Since, as mentioned above, $H$ in the original coordinates is given by Eq.~\eqref{eqn:hamiltonian_geodesic}, it follows that in these good coordinates it is given by the same equation, but with $m$ defined as a function of $\pi_\alpha$ and $\hat S^{\hat \alpha \hat \beta}$ through a variable which we call $p_a (X)$:
\begin{equation} \label{eqn:hamiltonian_spin}
  H(X) = -\sqrt{-g^{ab} p_a (X) p_b (X)},
\end{equation}
where we have defined $p_a (X)$ by
\begin{equation} \label{eqn:auxiliary_vars}
  p_a (X) \equiv \pi_a + \frac{1}{2} (\omega_{\hat \alpha \hat \beta})_a \hat S^{\hat \alpha \hat \beta}.
\end{equation}
What has changed is that the non-zero components of the Poisson bivector are now determined by~\cite{Tauber1988, Khriplovich1989}
\begin{subequations}
  \begin{align}
    \{x^\alpha, \pi_\beta\} &= \delta^\alpha{}_\beta, \\
    \{\hat S^{\hat \alpha \hat \beta}, \hat S^{\hat \gamma \hat \delta}\} &= 2 (\eta^{\hat \alpha[\hat \delta} \hat S^{\hat \gamma] \hat \beta} - \eta^{\hat \beta[\hat \delta} \hat S^{\hat \gamma] \hat \alpha}).
  \end{align}
\end{subequations}
Note that now $\Pi^{AB}$ no longer depends explicitly on $g_{ab}$.

With the background motion now fully presented, we can now consider the motion of the spinning particle under the gravitational self-force.
As shown in~\cite{Detweiler2002, Pound2012a, Pound2017}, for a non-spinning particle, there exists an ``effective'' metric $\breve g_{ab} [\gamma; \varepsilon]$ (defined in Sec.~\ref{sec:sf_fields} below) such that
\begin{itemize}

\item the worldline $\gamma$ and the variable $p_a$ are now functions of $\varepsilon$,

\item the metric $g_{ab}$ is replaced with the effective metric $\breve g_{ab} [\gamma; \varepsilon]$, and

\item all quantities which are constructed from the metric ($\nabla_a$ and $\dot \gamma^a$ due to its normalization) are all replaced with versions ($\breve \nabla_a [\gamma; \varepsilon]$ and $\dot{\breve \gamma}^a [\gamma; \varepsilon]$) which are all constructed from the effective metric.

\end{itemize}
This means that we can describe non-spinning self-force motion using the following pseudo-Hamiltonian:
\begin{equation} \label{eqn:geodesic_pseudo}
  H(X, \bar X) = -\breve m(X, \bar X),
\end{equation}
where $\breve m$ is defined in terms of the effective metric $\breve g_{ab}$, considered as a function of some arbitrary point in phase space through the trajectory (defined using the background Hamiltonian) that passes through $\bar X$.
That is, $\breve m$ is defined as in Eq.~\eqref{eqn:m}, but with
\begin{equation} \label{eqn:effective}
  \breve g_{ab} (\bar X; \varepsilon) \equiv \breve g_{ab} [\pi \circ \Upsilon_\cdot (\bar X); \varepsilon]
\end{equation}
replacing $g_{ab}$, where $\Upsilon_\cdot (\bar X)$ abstractly denotes the curve determined by considering $\Upsilon_\Delta (\bar X)$ as a function of $\Delta$.

For the case of spinning body motion, we assume\footnote{While the prescription above for non-spinning motion has been shown to hold, even to $O(\varepsilon^2)$, and \emph{regardless of the nature of the body} (that is, whether it is a black hole or some material body) by~\cite{Pound2012a, Pound2017}, using an effective metric defined in~\cite{Pound2012b}, the validity of the analogous results to linear order in spin are somewhat murkier (see the discussion in Sec.~II.B of~\cite{Mathews2021}).
  Notably, despite the fact that this prescription for generating the equations of motion is known to hold at $O(\varepsilon)$ for material bodies from~\cite{Harte2011}, for a particular choice of effective metric, it is unclear if it also holds for the effective metric in~\cite{Pound2012a, Pound2012b, Pound2017}.
  We thank Sam Upton for confirming that work on extending the results of~\cite{Pound2012a, Pound2012b, Pound2017} to the spinning case is still ongoing.
  See~\cite{Mathews2021} for arguments for why this prescription is plausible.
  For simplicity, we will assume that this prescription holds. \label{foot:caveat}} that a similar prescription applies: in addition to the points described above for non-spinning motion, the equations of motion are the same as those described above in Eq.~\eqref{eqn:mpd_reduced}, but with the following modifications:
\begin{itemize}

\item $S^{ab}$ is also a function of $\varepsilon$, and

\item the Riemann tensor $R^a{}_{bcd}$ is replaced by $\breve R^a{}_{bcd} [\gamma; \varepsilon]$, which is constructed from the effective metric.

\end{itemize}
From this, the pseudo-Hamiltonian is given by modifying Eq.~\eqref{eqn:hamiltonian_spin} as in Eq.~\eqref{eqn:geodesic_pseudo}:
\begin{equation}
  H(X, \bar X; \varepsilon) \equiv -\sqrt{-\breve g^{ab} (\bar X; \varepsilon) \breve p_a (X, \bar X; \varepsilon) \breve p_b (X, \bar X; \varepsilon)},
\end{equation}
where $\breve p_a$ is defined analogously to how it is defined in Eq.~\eqref{eqn:auxiliary_vars}, except by using a $(\breve e_{\hat \alpha})^a (\bar X; \varepsilon)$ and $\breve \omega_{\hat \alpha \hat \beta} (\bar X; \varepsilon)$ that are defined from $\breve g_{ab} (\bar X; \varepsilon)$ and its compatible covariant derivative, instead of $g_{ab}$, using Eqs.~\eqref{eqn:tetrad} and~\eqref{eqn:spin_connection}, respectively.
Note that $(e^{\hat \alpha})_a$ does not appear in Eq.~\eqref{eqn:auxiliary_vars}, and so $(\breve e^{\hat \alpha})_a (\bar X; \varepsilon)$ does not arise in this equation.
As mentioned above, since the components of $\Pi^{AB}$ in these coordinates are independent of $g_{ab}$, there is \emph{no} modification of $\Pi^{AB}$ in this gauge to an analogous $\breve \Pi^{AB}$.

\section{Field equations} \label{sec:fields}

Next, we review the perturbative field equations in the problem of spinning, self-forced motion.
We start by describing the notation which we will use in the rest of this section.
Suppose that one has a tensor field $\bs Q[\bs \Phi]$ which is a functional of some other tensor field $\Phi_{\ms A}$, whose indices we collectively denote with the composite index $\ms A$.
For example, one could consider the case where $\Phi_{ab}$ is the metric $g_{ab}$, in which case $\ms A$ represents $ab$.
The functional dependence of $\bs Q$ on $\bs \Phi$ is, in general, nonlinear, although we will assume that it is still \emph{local}.

Now, suppose that $\Phi_{\ms A}$ is a function of a single parameter $\lambda$.
It then follows that one can define, from $\bs Q$, a local, linear functional $\bs Q' (\lambda)$ by writing
\begin{equation}
  \bs Q' (\lambda) \left\{\frac{\partial \bs \Phi}{\partial \lambda}\right\} \equiv \frac{\partial \bs Q}{\partial \lambda},
\end{equation}
where we have dropped the functional dependence on $\Phi_{\ms A}$ on both sides, but we have explicitly included the dependence on $\lambda$.
When we set $\lambda = 0$, we write the partial derivative as a variation $\var$ [as in Eq.~\eqref{eqn:var}] and omit the dependence on $\lambda$ from $\bs Q'$:
\begin{equation}
  \bs Q' \{\var \bs \Phi\} \equiv \var \bs Q.
\end{equation}
We use curly braces in these equations to indicate the fact that $\bs Q' (\lambda)$ (or $\bs Q'$) is a \emph{local, linear functional} of $\partial \Phi_{\ms A}/\partial \lambda$ (or $\var \Phi_{\ms A}$).
Occasionally, it is useful to consider local, linear functionals of a single argument as operators, which can be accomplished by giving them an additional index (dual to that of their argument), and in an index-free context we denote their action on their argument with a $\cdot$.
That is, for such a functional $\bs{\mc Q}$ which we evaluate with some argument $\Phi_{\ms A}$, we can interchangeably write
\begin{equation}
  \bs{\mc Q} \{\bs \Phi\} \equiv \bs{\mc Q}^{\ms A} \Phi_{\ms A} \equiv \bs{\mc Q} \cdot \bs \Phi.
\end{equation}

While $\bs Q' (\lambda)$ reflects the linear dependence of $\bs Q$ on $\bs \Phi$, the quadratic dependence of $\bs Q$ on $\bs \Phi$ will also be occasionally be relevant.
This dependence can be recovered from the remaining dependence of $\bs Q' (\lambda)$ on $\Phi_{\ms A}$, which we have neglected in the equations above.
This can be seen as follows: for any local functional $\bs{\mc Q} (\lambda)$, which depends nonlinearly on $\Phi_{\ms A} (\lambda)$ and linearly on some $\phi_{\ms A} (\lambda)$, we take a derivative with respect to $\lambda$ to define $\bs{\mc Q}' (\lambda)$:
\begin{equation}
  \bs{\mc Q}' (\lambda) \left\{\frac{\partial \bs \Phi}{\partial \lambda}, \bs \phi (\lambda)\right\} \equiv \frac{\partial \bs{\mc Q} (\lambda) \{\bs \phi (\lambda)\}}{\partial \lambda} - \bs{\mc Q} (\lambda) \left\{\frac{\partial \bs \phi}{\partial \lambda}\right\}.
\end{equation}
Unlike above, we explicitly eliminate any derivatives of the arguments using the second term on the right-hand side; in other words, we are taking the derivative with respect to $\lambda$, with the argument of $\bs{\mc Q}$ \emph{fixed}.
In the case where we set $\lambda = 0$, this becomes
\begin{equation}
  \bs{\mc Q}' \{\var \bs \Phi, \bs \phi\} = \var \bs{\mc Q} \{\bs \phi\} - \bs{\mc Q} \{\var \bs \phi\}.
\end{equation}
We can now use this procedure to define $\bs Q''$ by
\begin{equation}
  \begin{split}
    \bs Q'' \{\var_1 \bs \Phi, \var_2 \bs \Phi\} &\equiv \var_1 \bs Q' \{\var_2 \bs \Phi\} - \bs Q' \{\var_1 \var_2 \bs \Phi\} \\
    &= \var_1 \var_2 \bs Q - \bs Q' \{\var_1 \var_2 \bs \Phi\},
  \end{split}
\end{equation}
where the second line makes it apparent that $\bs Q''$, which is essentially the ``quadratic part'' of $\bs Q$, is symmetric in its two arguments.
It follows from these definitions that
\begin{equation}
  \begin{split}
    \bs Q (\lambda) &= \bs Q + \lambda \bs Q' \{\var \bs \Phi\} + \frac{\lambda^2}{2} \left(\bs Q' \{\var^2 \bs \Phi\} + \bs Q'' \{\var \bs \Phi, \var \bs \Phi\}\right) \\
    &\hspace{1.1em}+ O(\lambda^3).
  \end{split}
\end{equation}

With this notation established, in the remainder of this section we first describe a local, bilinear current $\bs \omega$, the symplectic current, which is conserved when the source-free, perturbative equations of motion are satisfied.
We then turn to the discussion of the perturbative field equations from which the effective metric $\breve g_{ab} [\gamma; \varepsilon]$ is constructed in the first-order gravitational self-force, using the retarded, singular, and regular metric perturbations $h_{ab}$, $h^S_{ab}$, and $h^R_{ab}$, respectively.
We conclude with a discussion of an important property relating the stress-energy tensor from which these metric perturbations are constructed and the pseudo-Hamiltonian which appears in the self-force equations of motion.

\subsection{Symplectic currents}

As it is central for the definition of our flux-balance law, we start by introducing a current, the \emph{symplectic current}, which will be conserved when the vacuum equations of motion are satisfied.
The conservation of such a current in vacuum regions is crucial for allowing flux-balance laws to be promoted from being local to the particle to truly global flux-balance laws involving quantities at the boundaries of spacetime.

The definition of the symplectic current begins with considering the Lagrangian formulation of the theory in question; while in this paper we only consider general relativity, the discussion here can be generalized to a theory for an arbitrary field $\Phi_{\ms A}$.
Unlike usual treatments of the Lagrangian formulation, we adopt the approach of~\cite{Lee1990, Burnett1990} in working entirely \emph{locally}: instead of using an action, we consider only the variation of the integrand which appears in the action, which is naturally given by a \emph{Lagrangian four-form} $\bs L$.
This description has two advantages over using the usual action approach: first, one does not need to introduce a compensating boundary term in theories such as general relativity where the Lagrangian depends on second derivatives of the field.
Second, the boundary term arising in the variation, which is typically neglected at various points in the action approach, appears far more prominently.

Upon variation of the Lagrangian four-form, one finds that (subject to certain technical assumptions, such as covariance~\cite{Lee1990})
\begin{equation} \label{eqn:lagrangian_variation}
  \var \bs L = \bs E^{\ms A} \var \Phi_{\ms A} + \ud \bs \theta\{\var \bs \Phi\}.
\end{equation}
Here, the tensor-valued four-form $\bs E^{\ms A}$ vanishes when the equations of motion are satisfied, and the three-form $\bs \theta$ is the integrand of the usual boundary term which arises when varying an action.

The three-form $\bs \theta$, often called the \emph{symplectic potential}~\cite{Lee1990}, can then be used to define the \emph{symplectic current} $\bs \omega$ as an antisymmetric, multilinear functional:
\begin{equation}
  \bs \omega\{\bs \Phi^1, \bs \Phi^2\} \equiv \bs \theta' \{\bs \Phi^1, \bs \Phi^2\} - (\interchange{1}{2}).
\end{equation}
By taking a second variation of Eq.~\eqref{eqn:lagrangian_variation}, one can show that
\begin{equation}
  \ud \bs \omega\{\bs \Phi^1, \bs \Phi^2\} = \Phi^1_{\ms A} (\bs E')^{\ms A} \{\bs \Phi^2\} - (\interchange{1}{2}).
\end{equation}
That is, it is conserved when the linearized equations of motion hold.
This is the key property of the symplectic current which we will use throughout the remainder of this paper.

Note, however, that there is an issue with using the symplectic product in order to define a flux at infinity: it is a \emph{bilinear}, \emph{antisymmetric} current, which requires that we have two different field perturbations on the background spacetime, each satisfying the equations of motion, in order to construct a conserved current.
On the other hand, physically relevant fluxes are typically \emph{quadratic} in the field.
However, if we were to have an operator, which we denote by $\mc D_{\ms A}{}^{\ms B}$, which maps the space of solutions to the vacuum equations of motion to itself, we could resolve this issue by using such a map to define a quadratic current from a \emph{single} perturbation~\cite{Burnett1990}.
Such an operator is called a \emph{symmetry operator}~\cite{Miller1977}.

Explicitly, a symmetry operator is an operator $\mc D_{\ms A}{}^{\ms B}$ satisfying
\begin{equation}
  (\bs E')^{\ms A} \{\bs{\mc D} \cdot \var \bs \Phi\} = \widetilde{\mc D}^{\ms A}{}_{\ms B} (\bs E')^{\ms B} \{\var \bs \Phi\},
\end{equation}
for some other operator $\widetilde{\mc D}^{\ms A}{}_{\ms B}$.
In this sense, symmetry operators ``commute with (the perturbed) equations of motion, up to equations of motion''.
For any field theory, one can always construct a symmetry operator from a vector field $\xi^a$ as follows: consider the fields $\bs \psi$ from which the equations of motion $\bs E^{\ms A}$ are constructed.
In the case of gravity, this is simply the (dynamical) field $g_{ab}$, but for theories with non-dynamical fields (such as Klein-Gordon theory or electromagnetism on a fixed background spacetime) $\bs \psi$ would include both $\Phi_{\ms A}$ and the non-dynamical fields (such as the metric).
If this vector field satisfies $\lie_\xi \bs \psi = 0$, then $\lie_\xi$ will be a symmetry operator.
For gravity, or for theories with a fixed metric that are linearized off of a background where $\Phi_{\ms A} = 0$, this is simply the condition that $\lie_\xi g_{ab} = 0$, or that $\xi^a$ generates isometries.
In addition, however, for specific theories and backgrounds, there exist symmetry operators \emph{not} associated with isometries
\begin{itemize}

\item in the presence of a background, rank-two Killing tensor, for Klein-Gordon theory~\cite{Carter1977};

\item in the presence of a background, rank-two Killing-Yano tensor for Dirac (and Weyl) fermions~\cite{Carter1979, Kalnins1989} and electromagnetism~\cite{Kalnins1989, Kalnins1992, Grant2019}; and

\item for linearized gravity on the Kerr spacetime~\cite{Aksteiner2016, Grant2020, Green2022}, due to it being Petrov type D, and therefore possessing a rank-two Killing-Yano tensor~\cite{Walker1970}.

\end{itemize}
In this paper, we will, however, not be concerned with \emph{any} of these symmetry operators, and instead focus on a symmetry operator introduced in~\cite{Grant2022b} which arises \emph{only} in the case of solutions to the self-force equations of motion given in Sec.~\ref{sec:sf_fields} below.

Inserting any symmetry operator into the symplectic current, we can define a new current which satisfies
\begin{equation} \label{eqn:symplectic_symmetry}
  \begin{split}
    \ud \bs \omega\{\bs \Phi^1, \bs{\mc D} \cdot \bs \Phi^2\} &= \Phi^1_{\ms A} \widetilde{\mc D}^{\ms A}{}_{\ms B} (\bs E')^{\ms B} \{\bs \Phi^2\} \\
    &\hspace{1.1em}- (\bs E')^{\ms A} \{\bs \Phi^1\} \mc D_{\ms A}{}^{\ms B} \Phi^2_{\ms B}.
  \end{split}
\end{equation}
In particular, this current is now no longer antisymmetric in $\Phi^1_{\ms A}$ and $\Phi^2_{\ms A}$, and can therefore be used to define a quadratic current from a single perturbation $\Phi^1_{\ms A} = \Phi^2_{\ms A}$.
In the case where $\Phi^1_{\ms A}$ and $\Phi^2_{\ms A}$ are both vacuum solutions to the linearized field equations, it then follows that the right-hand side of Eq.~\eqref{eqn:symplectic_symmetry} vanishes, and so the current is conserved.

For concreteness, we now give the expression for this symplectic current in the relevant theory, general relativity.
The Einstein-Hilbert Lagrangian is given by\footnote{Note that the overall coefficient in front of the Einstein-Hilbert Lagrangian is arbitrary, so long as the stress-energy tensor, defined in terms of the variation of the matter Lagrangian, is such that $G^{ab} = 8\pi T^{ab}$ still holds.
  We adopt the choice in Eq.~\eqref{eqn:gr_lagrangian} such that Eq.~\eqref{eqn:gr_eom} takes a particularly simple form.}
\begin{equation} \label{eqn:gr_lagrangian}
  \bs L = -R \bs \epsilon,
\end{equation}
which implies that
\begin{equation} \label{eqn:gr_eom}
  \bs E^{ab} = G^{ab} \bs \epsilon.
\end{equation}
Moreover, the symplectic potential is given by
\begin{equation}
  \theta_{abc} \{\bs h\} = -2 (C')^{[d}{}_{ef} \{\bs h\} g^{e]f} \epsilon_{dabc},
\end{equation}
where $(C')^a{}_{bc} \{\bs h\}$ is the linearized connection coefficient constructed from the metric perturbation $h_{ab}$.
The symplectic current is then given by
\begin{equation}
  \begin{split}
    \omega_{abc} \{\bs h^1, \bs h^2\} = 2 \Big[&(C')^{[d}{}_{ef} \{\bs h^2\} P^{e]gfh}  h^1_{gh} \\
    &- (\interchange{1}{2})\Big] \epsilon_{dabc},
  \end{split}
\end{equation}
where
\begin{equation}
  P^{egfh} \equiv g^{(e|g} g^{|f)h} - \frac{1}{2} g^{ef} g^{gh}
\end{equation}
is the trace-reversal operator.

\subsection{Self-force} \label{sec:sf_fields}

As promised above in Sec.~\ref{sec:pole_dipole}, we now determine the effective metric $\breve g_{ab}$ which appears in the equations of motion.
First, consider the retarded solution $h_{ab}$ that is sourced by the worldline $\gamma$ by the following partial differential equation:
\begin{equation} \label{eqn:eom_retarded}
  (\bs E')^{ab} \{\bs h[\gamma]\} = 8\pi \bs T^{ab} [\gamma],
\end{equation}
where, from Eq.~\eqref{eqn:gr_eom} (and the fact that we are in a vacuum background spacetime), $(\bs E')^{ab}$ is a differential-form version of the linearized Einstein operator and $\bs T^{ab} [\gamma]$ is a distributional source that is given below in Eq.~\eqref{eqn:stress_energy}.
In particular, $h_{ab}$ is a vacuum solution off of $\gamma$.
In a convex normal neighborhood of $\gamma$, one can split the retarded field $h_{ab}$ into two fields, the \emph{singular field} $h^S_{ab}$, which is the part which is \emph{not} a vacuum solution, and in fact blows up, on $\gamma$:
\begin{equation} \label{eqn:eom_singular}
  (\bs E')^{ab} \{\bs h^S [\gamma]\} = 8\pi \bs T^{ab} [\gamma];
\end{equation}
and the \emph{regular field} $h^R_{ab}$, which remains a smooth vacuum solution, even on $\gamma$:
\begin{equation}
  (\bs E')^{ab} \{\bs h^R [\gamma]\} = 0.
\end{equation}
The effective metric is then defined by
\begin{equation}
  \breve g_{ab} [\gamma; \varepsilon] \equiv g_{ab} + \varepsilon h^R_{ab} [\gamma].
\end{equation}

While it is unimportant for the discussion in this paper, one can operationally compute these fields as follows: first, up to a given order in distance from the source, $h^S_{ab}$ has a known analytic form (given, for example, in~\cite{Pound2012b}).
After solving for the full retarded solution $h_{ab}$, one can then simply subtract off $h^S_{ab}$ in order to obtain $h^R_{ab}$.
Typically, this is only done at some truncated order in distance to the particle, but at sufficiently high order that the value of $h^R_{ab}$ and its first derivative can be recovered exactly.
At second order, this discussion becomes more complicated; see for example~\cite{Upton2021}.

Finally, we need to describe the differential-form version of the stress-energy tensor on the right-hand side of Eqs.~\eqref{eqn:eom_retarded} and~\eqref{eqn:eom_singular}.
In terms of the variables $p_a$ and $S^{ab}$, it is given by~\cite{Pound2012b}
\begin{equation}
  \begin{split}
    \bs T^{ab} [\gamma] = \int_{-\infty}^\infty \frac{\ud \tau'\; g^{a'd'} p_{d'}}{m(X')} \Bigg(&g^{b'c'} p_{c'} \bs \delta^{ab}{}_{a'b'} \\
    &+ S^{b'c'} \nabla_c \bs \delta^{(ab)c}{}_{a'b'c'}\Bigg),
  \end{split}
\end{equation}
where $x' \equiv \gamma(\tau')$ and $\delta_{abcd}{}^{\ms E}{}_{\ms E'}$ is a bitensor-valued four-form distribution (note that the four-form indices are at $x$, the same point as the first set of indices).
This distribution is defined by the following property: when integrated against a tensor $f_{\ms A}$ over some volume $V$, one finds
\begin{equation} \label{eqn:distribution}
  \int_V f_{\ms A} \bs \delta^{\ms A}{}_{\ms A'} = \begin{cases}
    f_{\ms A'} & x' \in V \\
    0 & x' \not \in V
  \end{cases}.
\end{equation}
for any collections of indices $\ms A$ and $\ms B$.
As such, this stress-energy tensor is supported entirely on $\gamma$.
Moreover, the distribution $\bs \delta^{\ms A}{}_{\ms A'}$ satisfies the property that
\begin{equation}
  \bs \delta^{\ms A \ms B}{}_{\ms A' \ms B'} = \bs \delta^{\ms B \ms A}{}_{\ms B' \ms A'}.
\end{equation}
Further, note that, upon taking a derivative $\nabla_{b'}$ of Eq.~\eqref{eqn:distribution}, we find
\begin{equation}
  \begin{split}
    \int_V f_{\ms A} \nabla_{b'} \bs \delta^{\ms A}{}_{\ms A'} &= \int_V (\nabla_b f_{\ms A}) \bs \delta^{b \ms A}{}_{b' \ms A'} \\
    &= -\int_V f_{\ms A} \nabla_b \bs \delta^{b \ms A}{}_{b' \ms A'},
  \end{split}
\end{equation}
and so
\begin{equation}
  \nabla_{b'} \bs \delta^{\ms A}{}_{\ms A'} = -\nabla_b \bs \delta^{b \ms A}{}_{b' \ms A'}.
\end{equation}
This implies that we can write
\begin{equation} \label{eqn:stress_energy}
  \begin{split}
    \bs T^{ab} [\gamma] = \int_{-\infty}^\infty \frac{\ud \tau'\; g^{a'd'} p_{d'}}{m(X')} \left(g^{b'c'} p_{c'} - S^{b'c'} \nabla_{c'}\right) \bs \delta^{(ab)}{}_{a'b'}.
  \end{split}
\end{equation}

Finally, we discuss a very useful property of the self-force in both the non-spinning and linear-in-spin cases, which we call \emph{Hamiltonian alignment}.
This is a relationship between the pseudo-Hamiltonian $H(\varepsilon)$ and the stress-energy tensor which appears in the self-force equations of motion in Eq.~\eqref{eqn:stress_energy}.
To introduce some notation, we define, for any tensor $\bs Q(X, \bar X; \varepsilon)$ which depends on $\bar X$ and $\varepsilon$ through $\varepsilon h^R_{ab} (\bar X)$, the tensors
\begin{equation} \label{eqn:partial}
  (\partial \bs Q)^{c_1 \cdots c_n ab} (X) \equiv \left.\frac{\partial \bs Q(X, \bar X; \varepsilon)}{\partial \nabla_{c_1} \cdots \nabla_{c_n} h^R_{ab} (\bar X)}\right|_{\bs h^R (\bar X) = 0}.
\end{equation}
The Hamiltonian alignment property is that, for \emph{some} constant $C_g$,
\begin{equation} \label{eqn:hamiltonian_alignment}
  \bs T^{a'b'} [\pi \circ \Upsilon_{\cdot} (X)] = C_g \int_{-\infty}^\infty \ud \tau'' \mc T^{a'b'} (X, \tau''),
\end{equation}
where, defining $X'' = \Upsilon_{\tau'' - \tau} (X)$ and $x'' = \pi(X'')$,
\begin{equation} \label{eqn:T}
  \begin{split}
    \mc T^{a'b'} (X, \tau'') = \sum_{n = 0}^\infty &(\partial H)^{c_1'' \cdots c_n'' a''b''} (X'') \\
    &\times \nabla_{c_1''} \cdots \nabla_{c_n''} \bs \delta^{a'b'}{}_{a''b''}.
  \end{split}
\end{equation}
Note that, while such a relationship is plausible from the generalized equivalence principle and the fact that both the stress-energy tensor and the background Hamiltonian must come from an action principle, we have not been able to prove this result more generally.
A version of this relationship also holds for the scalar field theory considered in~\cite{Grant2022b}, which (although not emphasized there) was key in the derivation of the flux-balance law in that paper.

In the case of a point particle, first note that
\begin{equation} \label{eqn:partial_m}
  (\partial m)^{ab} (X) = \frac{g^{ac} g^{bd} p_c p_d}{2 m(X)}.
\end{equation}
Using Eq.~\eqref{eqn:T}, and comparing with Eq.~\eqref{eqn:stress_energy}, we find that the Hamiltonian alignment condition holds, with $C_g = -1/2$.
There are no terms in Eq.~\eqref{eqn:T} beyond $n = 0$, since the pseudo-Hamiltonian in this case does not depend upon derivatives of the metric perturbation.
To linear order in spin, we have that
\begin{equation} \label{eqn:partial_H}
  \begin{split}
    (\partial H)^{ab} (X) = \frac{g^{cd} p_c (X)}{2 m(X)} \Big[&-\delta^a{}_d g^{be} p_e (X) \\
    &+ (\partial \omega_{\hat \alpha \hat \beta})_d{}^{ab} \hat S^{\hat \alpha \hat \beta}\Big].
  \end{split}
\end{equation}
However, from Eq.~\eqref{eqn:tetrad}, we find that
\begin{equation}
  (\partial e_{\hat \alpha})^{abc} = -\frac{1}{2} g^{a(b} (e_{\hat \alpha})^{c)},
\end{equation}
\{compare Eq.~(39b) of~\cite{Blanco2023}\}, and so
\begin{equation} \label{eqn:partial_omega}
  \begin{split}
    (\partial \omega_{\hat \alpha \hat \beta})_a{}^{bc} &= (e_{\hat \alpha})^{(b} \nabla_a (e_{\hat \beta})^{c)} - \frac{1}{2} g_{de} g^{d(b} (e_{\hat \alpha})^{c)} \nabla_a (e_{\hat \beta})^e \\
    &\hspace{1.1em}- \frac{1}{2} g_{de} (e_{\hat \alpha})^d \nabla_a [g^{e(b} (e_{\hat \beta})^{c)}] \\
    &= 0,
  \end{split}
\end{equation}
where in the first equality we have used the fact that $\breve \nabla_a$ only depends on $\nabla_a h^R_{bc}$ to linear order.
As such, the second term in Eq.~\eqref{eqn:partial_H} doesn't contribute, and so
\begin{equation} \label{eqn:partial_H2}
  (\partial H)^{ab} (X) = -\frac{g^{ac} g^{bd} p_c (X) p_d (X)}{2 m(X)}.
\end{equation}

Now, we consider $(\partial H)^{abc} (X)$, which is given by
\begin{equation}
  (\partial H)^{abc} (X) = \frac{g^{cd} p_c (\partial \omega_{\hat \alpha \hat \beta})_d{}^{abc} \hat S^{\hat \alpha \hat \beta}}{2 m(X)}.
\end{equation}
We have that
\begin{equation} \label{eqn:partial_omega3}
  \begin{split}
    (\partial \omega_{\hat \alpha \hat \beta})_a{}^{bcd} = g_{ef} (e_{\hat \alpha})^e \bigg[&(\partial C)^f{}_{ga}{}^{bcd} (e_{\hat \beta})^g \\
    &- \frac{1}{2} \delta_a{}^b g^{f(c} (e_{\hat \beta})^{d)}\bigg],
  \end{split}
\end{equation}
where
\begin{equation} \label{eqn:partial_C}
  \begin{split}
    (\partial C)^e{}_{fa}{}^{bcd} &= \frac{1}{2} g^{eg} \left(2 \delta^b{}_{(f} \delta^{(c}{}_{a)} \delta^{d)}{}_g - \delta^b{}_g \delta^{(c}{}_f \delta^{d)}{}_a\right) \\
    &= \frac{1}{2} g^{eg} \left(\delta^b{}_a \delta^{(c}{}_f \delta^{d)}{}_g + 2 \delta^b{}_{[f} \delta^{(c}{}_{g]} \delta^{d)}{}_a\right)
  \end{split}
\end{equation}
reflects the dependence of the connection coefficients on $\nabla_b h^R{}_{cd}$.
The first term on the second line of this equation cancels with the term on the second line of Eq.~\eqref{eqn:partial_omega3}, and so we find that
\begin{equation}
  (\partial \omega_{\hat \alpha \hat \beta})_a{}^{bcd} = -(e_{[\hat \alpha})^b (e_{\hat \beta]})^{(c} \delta^{d)}{}_a.
\end{equation}
This implies that
\begin{equation} \label{eqn:partial_H3}
  (\partial H)^{abc} (X) = \frac{g^{d(b} p_d (X) S^{c)a}}{2 m(X)}.
\end{equation}
By comparing Eqs.~\eqref{eqn:partial_H2} and~\eqref{eqn:partial_H3} to Eq.~\eqref{eqn:stress_energy} and~\eqref{eqn:T}, we see that, even in the linear-in-spin case, the Hamiltonian alignment condition holds, with $C_g = -1/2$.
We do not need to go beyond $n = 1$ in Eq.~\eqref{eqn:partial} because there are no terms in the pseudo-Hamiltonian which depend on more than one derivative of the metric perturbation.

It is at this point that we can comment on the reason why it is necessary for us to work on the 14-dimensional, ``fictitious'' phase space where the spin-supplementary condition $d^a = 0$ is not required to hold.
An alternative approach would have been to use the Hamiltonian in Eq.~\eqref{eqn:hamiltonian_spin} in coordinates where one has already restricted to the surface where $d^a = 0$.
However, in writing these coordinates in terms of the well-behaved coordinates in Eq.~\eqref{eqn:good_coords}, one needs to explicitly introduce factors of the metric, which in passing to the self-force pseudo-Hamiltonian would introduce an additional dependence on $h^R_{ab} (\bar X)$.
This would spoil the Hamiltonian alignment condition, and so for simplicity we work on the larger, 14-dimensional phase space.

\section{General flux-balance laws} \label{sec:flux_balance}

We now discuss our approach to flux-balance laws in general spacetimes.
First, as discussed above, any metric perturbation $h^{\cdots}_{ab}$ in this problem (for example, the full retarded field $h_{ab}$), depends on a worldline $\gamma$ [for example, via Eq.~\eqref{eqn:eom_retarded}].
As such, as in Eq.~\eqref{eqn:effective}, it can be considered as a function of a point $X$ which lies along a trajectory $\Gamma$ through phase space at some fixed time $\tau$, via the Hamiltonian flow map $\Upsilon$:
\begin{equation}
  h^{\cdots}_{a'b'} (X) \equiv h^{\cdots}_{a'b'} [\pi \circ \Upsilon_\cdot (X)].
\end{equation}
As in~\cite{Grant2022b}, we then consider the operation given by taking a derivative with respect to $X$, which (as before) we denote by $\nabla_A$.
Note that derivatives of $h^{\cdots}_{a'b'} (X)$ with respect to $X$ and $x'$ act independently since, considered as a bitensor at $X$ and $x'$, $h^{\cdots}_{a'b'} (X)$ does not impose any relationship through its definition (unlike, say, $\Upsilon^{A'}{}_A$).
As such, $\nabla_A$ and $\nabla_{a'}$ commute, and moreover $\nabla_A g_{a'b'} = 0$ (the background metric is not a function of $X$), so $\nabla_A$ is a symmetry operator.

\begin{figure}
  \includegraphics[width=\linewidth]{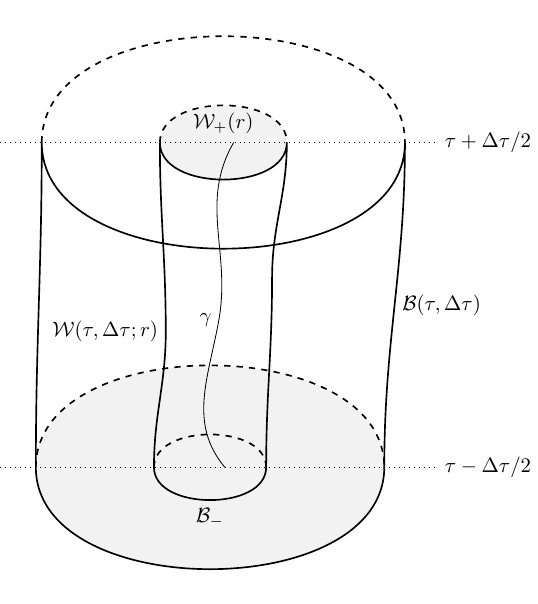}
  \caption{The various surfaces surrounding the worldline $\gamma$ on which the flux-balance laws are integrated: the inner worldtube $\mc W(\tau, \Delta \tau; r)$, the outer boundary $\mc B(\tau, \Delta \tau)$, and the two pairs of endcaps, $\mc W_\pm (r)$ and $\mc B_\pm$.
    For clarity, we only label (and shade) two of the endcaps, $\mc W_+ (r)$ (at the top) and $\mc B_-$ (at the bottom).
    \label{fig:integration}}
\end{figure}

Using this symmetry operator, we can define a flux integral over some hypersurface, which we generically denote by $\Sigma(\tau, \Delta \tau)$.
We assume that this hypersurface is a tube surrounding the worldline $\gamma$, with boundaries which we denote by $\partial \Sigma_\pm$.
These surfaces $\partial \Sigma_\pm$ are also the boundaries of hypersurfaces $\Sigma_\pm$ which intersect $\gamma$ at $\gamma(\tau_\pm)$, where $\tau \pm \equiv \tau \pm \Delta \tau/2$.
The two hypersurfaces we will be integrating our fluxes over will consist of an outer boundary $\mc B(\tau, \Delta \tau)$, which we will later assume can be extended down to the event horizon $\scrh$ and out to null infinity $\scri$, and a worldtube $\mc W(\tau, \Delta \tau; r)$ of some proper radius $r$.
Note that $\mc W(\tau, \Delta \tau; r)$ needs to be within the convex normal neighborhood of $\gamma$.
As a visual aid, see Fig.~\ref{fig:integration}.

In terms of a general surface $\Sigma(\tau, \Delta \tau)$, our flux integral, as a function of two metric perturbations $h^1_{ab}$ and $h^2_{ab}$, will be given by\footnote{Note that, in this equation, the operator $\nabla_A$ appears inside the local, bilinear functional $\bs \omega$, which is then integrated.
  The validity of a free index inside the functional is due to (bi)linearity and locality, and the validity inside of an integral is due to the fact that it is an index at a point which is \emph{not} being integrated over.}
\begin{equation} \label{eqn:flux_def}
  \mc F^\Sigma_A \{\bs h^1, \bs h^2\} \equiv \lim_{\Delta \tau \to \infty} \frac{1}{8\pi C_g \Delta \tau} \int_{\Sigma(\tau, \Delta \tau)} \bs \omega\{\bs h^1, \nabla_A \bs h^2\},
\end{equation}
where $X$, the point at which $\nabla_A$ is applied, is given by $\Gamma(\tau)$; the integrand is evaluated at some point $x'$ on the surface $\Sigma(\tau, \Delta \tau)$; and $C_g$ is the constant (in gravity equal to $-1/2$) appearing in the Hamiltonian alignment condition in Eq.~\eqref{eqn:hamiltonian_alignment}.

Since $h_{ab}$, the retarded field, is a vacuum solution in the region between $\mc B$ and $\mc W$, we find
\begin{equation} \label{eqn:flux_W}
  \mc F^{\mc B}_A \{\bs h, \bs h\} = \mc F^{\mc W(r)}_A \{\bs h, \bs h\}
\end{equation}
for any (well-defined, that is, sufficiently small) proper radius $r$.
This follows by Stokes' theorem; note that, in general, there will be an integral over the surfaces $\mc B_\pm - \mc W_\pm (r)$, but as this region is finite, and we divide by $\Delta \tau$ and take the limit $\Delta \tau \to \infty$ in the definition of the fluxes, such a term goes away, and only integrals over $\mc B(\tau, \Delta \tau)$ and $\mc W(\tau, \Delta \tau; r)$ remain.\footnote{Note that this relies on the assumption that $h_{a'b'} (X)$ and $\nabla_A h_{a'b'} (X)$ both remain finite as one takes $\tau' - \tau \to \pm \infty$; as we will see below in Sec.~\ref{sec:geodesic_kerr}, this is only true for \emph{certain components} of $\nabla_A h_{a'b'} (X)$.
  However, when $\nabla_A h_{a'b'} (X)$ blows up, the flux $\mc F^{\mc B}_A \{\bs h, \bs h\}$ will \emph{also} diverge, and so this is a case in which we cannot use these flux-balance laws anyways.}
In the remainder of the paper, we will not discuss these ``endcap'' contributions which would otherwise appear from applications of Stokes' theorem, as they all vanish for this same reason.

At this point, by the bilinearity of our current, we can now split the right-hand side into pieces which come from $h^R_{ab}$ and $h^S_{ab}$.
As we will show in the next two sections, the following result holds:
\begin{equation} \label{eqn:quasilocal_result}
  \begin{split}
    \left\<\var \dot \Gamma^A (\tau')\right\>_{\tau'} = \Pi^{AB} \bigg\{\frac{1}{2} \mc F^{\mc B}_B \{\bs h, \bs h\} + \nabla_B \<\mc H(X')\>_{\tau'}\bigg\},
  \end{split}
\end{equation}
where for brevity we write $X' \equiv \Upsilon_{\tau' - \tau} (X)$, and $\mc H$ is the Hamiltonian for the conservative dynamics described in detail in~\cite{Blanco2022, Blanco2023, Blanco2024}, and discussed in more detail in Sec.~\ref{sec:conservative} below.

\subsection{Local flux-balance laws}

We first consider local flux-balance laws at the worldtube $\mc W(\tau, \Delta \tau; r)$ around the particle.
We start with the fact that the chain rule, together with Eqs.~\eqref{eqn:var} and~\eqref{eqn:partial}, implies that
\begin{equation} \label{eqn:var_split}
  \var H(X', \bar X) = \sum_{n = 0}^\infty (\partial H)^{c_1' \cdots c_n' a'b'} (X') \nabla_{c_1'} \cdots \nabla_{c_n'} h^R_{a'b'} (\bar X).
\end{equation}
Inserting Eq.~\eqref{eqn:var_split} into Eq.~\eqref{eqn:avg_vardot}, we find that
\begin{widetext}
\begin{equation} \label{eqn:avg_vardot_split}
  \left\<\var \dot \Gamma^A (\tau')\right\>_{\tau'} = \Pi^{AB} \left\<\left[\nabla_B \sum_{n = 0}^\infty (\partial H)^{c_1' \cdots c_n' a'b'} \{\Upsilon_{\tau' - \tau} (X)\} \nabla_{c_1'} \cdots \nabla_{c_n'} h^R_{a'b'} (\bar X)\right]_{\bar X \to X}\right\>_{\tau'}.
\end{equation}
We then replace the factors of $h^R_{a'b'}$ with integrals over delta functions, so that, by using Eq.~\eqref{eqn:T},
\begin{equation} \label{eqn:avg_vardot_deltas}
  \left\<\var \dot \Gamma^A (\tau')\right\>_{\tau'} = \Pi^{AB} \left\<\left[\nabla_B \int_{\mc V(\tau, \infty; r)} h^R_{a''b''} (\bar X) \bs{\mc T}^{a''b''} (X, \tau')\right]_{\bar X \to X}\right\>_{\tau'},
\end{equation}
\end{widetext}
where we have used the fact that the inner covariant derivatives acting on $h^R_{a'b'} (\bar X)$ in Eq.~\eqref{eqn:avg_vardot_split} are now applied to the delta functions, and where $\mc V(\tau, \Delta \tau; r)$ is the volume containing the worldline $\gamma$ such that its boundary is the worldtube:
\begin{equation}
  \partial \mc V(\tau, \Delta \tau; r) = \mc W(\tau, \Delta \tau; r) \cup \mc W_+ (r) \cup \mc W_- (r)
\end{equation}
[and similarly $\partial \mc V(\tau, \infty; r) = \mc W(\tau, \infty; r)$].
What this has done is ``factor'' the dependence on $\bar X$ onto something which is now clearly is independent of $X$, and so (after interchanging the order of the integrals) we can simply \emph{set} $\bar X = X$.
In interchanging the order of the integrals, we can also switch the bounds on the $\tau'$ integral from $\tau_\pm$ to $\pm \infty$ and from the volume $\mc V(\tau, \infty; r)$ to $\mc V(\tau, \Delta \tau; r)$:
\begin{equation} \label{eqn:avg_vardot_factored}
  \begin{split}
    \left\<\var \dot \Gamma^A (\tau')\right\>_{\tau'} = \lim_{\Delta \tau \to \infty} \frac{\Pi^{AB}}{\Delta \tau} &\int_{\mc V(\tau, \Delta \tau; r)} h^R_{a''b''} (X) \\
    &\times \nabla_B \int_{-\infty}^\infty \ud \tau'\; \bs{\mc T}^{a''b''} (X, \tau').
  \end{split}
\end{equation}

So far, the discussion has been somewhat theory-agnostic.
However, in the case where the Hamiltonian alignment condition in Eq.~\eqref{eqn:hamiltonian_alignment} holds, we can now write
\begin{equation} \label{eqn:avg_vardot_T}
  \left\<\var \dot \Gamma^A (\tau')\right\>_{\tau'} = \lim_{\Delta \tau \to \infty} \frac{\Pi^{AB}}{C_g \Delta \tau} \int_{\mc V(\tau, \Delta \tau; r)} h^R_{a'b'} \nabla_B \bs T^{a'b'}, \vspace{\belowdisplayskip}
\end{equation}
where (for brevity) we drop the dependence on $X$ on the right-hand side.
We now use Stokes' theorem and Eq.~\eqref{eqn:symplectic_symmetry}, which implies that
\begin{equation} \label{eqn:flux_RS}
  \left\<\var \dot \Gamma^A (\tau')\right\>_{\tau'} = \Pi^{AB} \mc F^{\mc W(r)}_B \{\bs h^R, \bs h^S\}. \vspace{\belowdisplayskip}
\end{equation}
This is the same sort of flux-balance law which was originally proven in~\cite{Grant2022b}: as it is constructed from the regular and singular fields, it is only applicable near the worldline.

Now, by bilinearity, there will be a contribution to the flux term in Eq.~\eqref{eqn:flux_W} coming from the same term that appears on the right-hand side of Eq.~\eqref{eqn:flux_RS}, except with $h^R_{ab}$ and $h^S_{ab}$ switched.
We now compute this contribution.
First, note that Eq.~\eqref{eqn:symplectic_symmetry} implies that
\begin{widetext}
\begin{equation}
  \mc F^{\mc W(r)}_A \{\bs h^S, \bs h^R\} = -\lim_{\Delta \tau \to \infty} \frac{1}{C_g \Delta \tau} \int_{\mc V(\tau, \Delta \tau; r)} \bs T^{a''b''} \nabla_A h^R_{a''b''}.
\end{equation}
By using the Hamiltonian alignment condition and reversing the factorization above, we can write $\nabla_A h^R_{a''b''} (X)$ as $[\nabla_{\bar A} h^R_{a''b''} (\bar X)]_{\bar X \to X}$, and then interchange the order of integrals as before, and we can undo the remaining steps that appeared in deriving Eq.~\eqref{eqn:flux_RS}, obtaining
\begin{equation} \label{eqn:SR_H}
  \begin{split}
    \mc F^{\mc W(r)}_A \{\bs h^S, \bs h^R\} &= -\left\<\left[\nabla_{\bar A} \int_{\mc V(\tau, \infty; r)} h^R_{a''b''} (\bar X) \bs{\mc T}^{a''b''} (X, \tau')\right]_{\bar X \to X}\right\>_{\tau'} \\
    &= -\left\<\left[\nabla_{\bar A} \var H\{\Upsilon_{\tau' - \tau} (X), \bar X\}\right]_{\bar X \to X}\right\>_{\tau'}.
  \end{split}
\end{equation}
\end{widetext}

Now, we finally use Synge's rule, which states that, for any bitensor $\bs Q(X, X')$, the following coincidence limits are related~\cite{Synge1960, Poisson2011}:
\begin{equation}
  \begin{split}
    \nabla_A [\bs Q(X, \bar X)]_{\bar X \to X} &= [\nabla_A \bs Q(X, \bar X)]_{\bar X \to X} \\
    &\hspace{1.1em}+ [\nabla_{\bar A} \bs Q(X, \bar X)]_{\bar X \to X}.
  \end{split}
\end{equation}
As such, we find that [when combined with Eq.~\eqref{eqn:avg_vardot}] Eq.~\eqref{eqn:SR_H} becomes
\begin{equation}
  \begin{split}
    \left\<\var \dot \Gamma^A (\tau')\right\>_{\tau'} = \Pi^{AB} \bigg\{&\nabla_B \<\var H[\Upsilon_{\tau' - \tau} (X), X]\> \\
    &+ \mc F^{\mc W(r)}_B \{\bs h^S, \bs h^R\}\bigg\}.
  \end{split}
\end{equation}
In combination with Eq.~\eqref{eqn:flux_RS}, together with the fact that
\begin{equation}
  \mc F^{\mc W(r)}_A \{\bs h^R, \bs h^R\} = 0
\end{equation}
[by Stokes' theorem, Eq.~\eqref{eqn:symplectic_symmetry}, and the fact that $h^R_{ab}$ is a vacuum solution \emph{everywhere}], it follows that
\begin{equation} \label{eqn:almost_result}
  \begin{split}
    \left\<\var \dot \Gamma^A (\tau')\right\>_{\tau'} = \frac{1}{2} \Pi^{AB} \bigg\{&\mc F^{\mc W(r)}_B \{\bs h, \bs h\} - I^{SS} (r) \\
    &+ \nabla_B \<\var H[\Upsilon_{\tau' - \tau} (X), X]\>_{\tau'}\bigg\},
  \end{split}
\end{equation}
where
\begin{equation}
  I^{SS}_A (r) \equiv \mc F^{\mc W(r)}_A \{\bs h^S, \bs h^S\}.
\end{equation}

\subsection{Conservative Hamiltonian} \label{sec:conservative}

We now relate the third term on the right-hand side of Eq.~\eqref{eqn:almost_result} to the conservative Hamiltonian which appears in~\cite{Blanco2022, Blanco2023, Blanco2024}.
First, we have that
\begin{equation} \label{eqn:msH}
  \begin{split}
    \var H[\Upsilon_{\tau' - \tau} (X), X] &= \var H[\Upsilon_{\tau' - \tau} (X), \Upsilon_{\tau' - \tau} (X)] \\
    &\equiv 2 \ms H[\Upsilon_{\tau' - \tau} (X)],
  \end{split}
\end{equation}
by the fact that $H(\varepsilon)$ does not depend on where on the trajectory its second argument lies.
We call $\ms H$, which is a true Hamiltonian (in the sense that it only depends on one point in phase space) the ``coincidence Hamiltonian''.
An argument similar to that which is used to derive Eq.~\eqref{eqn:avg_vardot_T} then shows that the average of the coincidence Hamiltonian is given by
\begin{equation}
  \left\<\ms H[\Upsilon_{\tau' - \tau} (X)]\right\>_{\tau'} = \lim_{\Delta \tau \to \infty} \frac{1}{2 C_g \Delta \tau} \int_{\mc V(\tau, \Delta \tau; r)} h^R_{a'b'} \bs T^{a'b'},
\end{equation}
where for brevity we have dropped the dependence on $X$ in each term of the integral on the right-hand side.

We now consider how the regular field is sourced by the stress-energy tensor: it is given in terms of an integral against a ``regular two-point function'' $G^R_{a'b'a''b''}$~\cite{Detweiler2002}:
\begin{equation}
  \begin{split}
    h^R_{a'b'} (X) &= \int_{\mc V(\tau, \infty; r)} G^R_{a'b'a''b''} \bs T^{a''b''} (X) \\
    &= \int_{\mc V(\tau, \Delta \tau; r)} G^R_{a'b'a''b''} \bs T^{a''b''} (X) \\
    &\hspace{1.1em}+ R^R_{a'b'} (X, \Delta \tau),
  \end{split}
\end{equation}
where $R^R_{a'b'} (X, \Delta \tau)$ is a ``remainder'' term that must go to zero in the limit $\Delta \tau \to \infty$.
As such, we find that
\begin{widetext}
\begin{equation} \label{eqn:symmetric_integrand}
  \left\<\ms H[\Upsilon_{\tau' - \tau} (X)]\right\>_{\tau'} = \lim_{\Delta \tau \to \infty} \frac{1}{2 C_g \Delta \tau} \int_{\mc V(\tau, \Delta \tau; r)} \int_{\mc V(\tau, \Delta \tau; r)} G^R_{a'b'a''b''} \bs T^{a'b'} (X) \bs T^{a''b''} (X).
\end{equation}
\end{widetext}
Since the integrand is symmetric in the two copies of the stress-energy tensor, it must therefore follow that we can replace $G^R_{a'b'a''b''}$ with
\begin{equation}
  G^{\rm symm.}_{a'b'a''b''} = \frac{1}{2} \left(G^R_{a'b'a''b''} + G^R_{a''b''a'b'}\right).
\end{equation}
From this two-point function, we can define a ``symmetric'' part of the regular metric perturbation $h^R_{ab}$ by
\begin{equation}
  \begin{split}
    h^{\rm symm.}_{a'b'} (X) &\equiv \int_{\mc V(\tau, \infty; r)} G^{\rm symm.}_{a'b'a''b''} \bs T^{a''b''} (X) \\
    &= \int_{\mc V(\tau, \Delta \tau; r)} G^{\rm symm.}_{a'b'a''b''} \bs T^{a''b''} (X) \\
    &\hspace{1.1em}+ R^{\rm symm.}_{a'b'} (X, \Delta \tau),
  \end{split}
\end{equation}
for a similar remainder $R^{\rm symm.}_{a'b'} (X, \Delta \tau)$.
As such, we can replace $h^R_{ab}$ with $h^{\rm symm.}_{ab}$ in Eq.~\eqref{eqn:msH} directly, and so
\begin{equation} \label{eqn:conservative_hamiltonian}
  \left\<\var H[\Upsilon_{\tau' - \tau} (X), X]\right\>_{\tau'} = 2 \left\<\mc H[\Upsilon_{\tau' - \tau} (X)]\right\>_{\tau'},
\end{equation}
where $\mc H(X)$ is given by $\ms H(X)$, but using $h^{\rm symm.}_{ab}$ instead of $h^R_{ab}$.
As discussed in~\cite{Blanco2022, Blanco2023, Blanco2024}, this is the true Hamiltonian which appears in the conservative dynamics at first order, and so we call it the ``conservative Hamiltonian''.

\subsection{Vanishing of the ``divergent'' contribution}

To complete the proof, we finally show that
\begin{equation} \label{eqn:divergence}
  I^{SS}_A (r) = 0.
\end{equation}
To do so, we start by noting that $I^{SS}_A (r)$ is independent of $r$, by Stokes' theorem, Eq.~\eqref{eqn:symplectic_symmetry}, and the fact that $h^S_{ab}$ is a vacuum solution off of $\gamma$.
In particular, this means that it does not diverge as $r \to 0$, as one would na\"ively expect due to the fact that $h^S_{ab}$ blows up on $\gamma$.
We therefore formally evaluate $I^{SS}_A (r)$ as a power series in $r$, and look only at the term which is constant in $r$, which by the above argument can be the only contribution.

Evaluating $I^{SS}_A (r)$ involves the computation of an integral over a sphere of radius $r$ around $\gamma$.
As such, determining the value of $I^{SS}_A (r)$ can be performed by investigating the structure of the integrand as a function of the normal vector $n_a \equiv \nabla_a r$: if it contains an \emph{odd} number of factors of $n_a$, then it must vanish.

To show that this is the case, we consider the parity structure of the various tensors which appear in this integrand: expanding such a tensor $\bs Q$ in powers of $r$, we have that (for some integer $N$)
\begin{equation}
  \bs Q = \sum_{n = N}^\infty r^n \bs Q_n.
\end{equation}
If the tensor $\bs Q_n$ contains an even number of factors of $n_a$ if $n$ is even and an odd number of factors if $n$ is odd, then $\bs Q$ is said to have \emph{even} parity.
Similarly, if $\bs Q_n$ contains an odd number of factors of $n_a$ if $n$ is even and an even number of factors if $n$ is odd, then $\bs Q$ is said to have \emph{odd} parity.
If either of these two cases (even and odd parity) hold, then $\bs Q$ has \emph{definite} parity, and otherwise \emph{indefinite} parity; fortunately, all tensors which appear in the integrand of $I^{SS}_A (r)$ have definite parity.

As examples, any analytic function near $\gamma$ \emph{must} have even parity, from the Taylor expansion near $\gamma$; this is the case for the background metric, for example.
However, $h^S_{ab}$, as can be seen from Eqs.~(98-100) and~(124) of~\cite{Pound2012b}, has odd parity in Lorenz gauge.\footnote{Note that this particular part of the calculation is gauge-dependent: in the ``highly regular'' gauges of~\cite{Upton2021}, the parity structure is quite different, and in the usual ``radiation'' gauges used near null infinity and the horizon there are singularities which appear near the particle~\cite{Barack2001}.
  As we do not need to \emph{compute} the metric perturbation except for asymptotically, we are free to merely specify abstract properties for the gauge in which we are working: near the particle, it should approach the Lorenz gauge, and near null infinity and the horizon, it should approach radiation gauge.}
Similarly, the volume element $\bs \epsilon^{\mc W}$ on $\mc W(\tau, \Delta \tau; r)$ has an explicit factor of $n_a$ when compared to the spacetime volume element $\bs \epsilon$:\footnote{The plus sign here is due to the fact that $r$ increases as one approaches the surface $\mc W(\tau, \Delta \tau; r)$ from the inside; see the discussion in Appendix B.2 of~\cite{Wald1984} to see why this is the orientation that is compatible with Stokes' theorem.}
\begin{equation}
  \bs \epsilon = \bs n \wedge \bs \epsilon^{\mc W},
\end{equation}
and so when one pulls $\bs \omega$ back to $\mc W(\tau, \Delta \tau; r)$, the integrand has an explicit factor of $\bs n$ which is introduced.
The product of two even (or two odd) parity tensors has even parity, and the product of an odd and even parity tensor has odd parity.
Similarly, the covariant derivative $\nabla_a$ compatible with the background metric $g_{ab}$ must also be even, in the sense that it maps even tensors to even tensors and odd tensors to odd tensors; moreover, the symmetry operator $\nabla_A$ \emph{also} has this property.

From the examples given above, it therefore follows that the integrand of $I^{SS}_A (r)$ must be odd, as it is constructed from two copies of $h^S_{ab}$ and the factor of $n_a$ from pulling back the volume element to $\mc W(\tau, \Delta \tau; r)$, with everything else that appears being even.
Since we are evaluating this integrand at an even power in its Taylor series expansion (zero), it follows that $I^{SS}_A (r)$ is of the form of an integral over the sphere of constant $r$ of an odd number of factors of $n_a$, which vanishes.
This proves Eq.~\eqref{eqn:divergence}, which [along with Eq.~\eqref{eqn:conservative_hamiltonian}] shows that Eq.~\eqref{eqn:almost_result} becomes Eq.~\eqref{eqn:quasilocal_result}.

\section{Flux-balance laws in the Kerr spacetime} \label{sec:kerr}

We now consider the application of Eq.~\eqref{eqn:quasilocal_result} to the Kerr spacetime.
First, note that we can now take $\mc B$ to null infinity and the horizon, since for any two surfaces $\mc B$ and $\mc B'$ at finite distance from the particle, the contribution from the endcap integrals over $\mc B_\pm' - \mc B_\pm$ vanishes in the limit $\Delta \tau \to \infty$.
In particular, note that we are making the explicit choice of taking the limit where $\mc B(\tau; \Delta \tau)$ goes to the horizon and null infinity \emph{after} taking the limit $\Delta \tau \to \infty$.
Note that while $\gamma$, the background geodesic, will not puncture $\mc B$ (assuming a bound orbit), the exact worldline $\gamma(\varepsilon)$ will.

For concreteness, we now define a coordinate $w$ such that
\begin{equation}
  w(\tau) = t(\tau) + h(r),
\end{equation}
where $t$ and $r$ are in the usual Boyer-Lindquist coordinates, and $t(\tau)$ indicates the value of $t$ such that the background worldline $\gamma$ has coordinate time $t$ at proper time $\tau$.
We moreover suppose that $h(r)$ has the property that
\begin{equation}
  h(r) = \begin{cases}
    r^* & r \to r_H \\
    -r^* & r \to \infty
  \end{cases},
\end{equation}
where $r_H$ is the location of the horizon and $r^*$ is the tortoise coordinate in Kerr, such that $w$ becomes $v$ (ingoing Eddington-Finkelstein time) at the horizon and $u$ (retarded time) at null infinity.
With $w$ defined, we define our outer region $\mc B$ such that $\mc B_\pm$ are surfaces of constant $w(\tau_\pm)$.

Next, note that the retarded metric perturbation which we are considering vanishes on the past horizon and on past null infinity.
This is just a consequence of the fact that it is a field with retarded boundary conditions (that is, no incoming radiation at null infinity, and no outgoing radiation at the horizon).
Our integral over $\mc B$ then becomes one over the future horizon and future null infinity.
However, we need to write the flux integral in Eq.~\eqref{eqn:flux_def} in terms of $w$, which is a good coordinate at these surfaces: to do so, note that
\begin{equation}
  \Delta w \equiv w(\tau_+) - w(\tau_-) = \int_{\tau_-}^{\tau_+} z(\tau') \ud \tau',
\end{equation}
where $z \equiv \ud t/\ud \tau$ is the redshift for the background worldline.
As such, we recover Eq.~\eqref{eqn:result}, where we define the following flux term $\mc F_A$:
\begin{equation} \label{eqn:boundary_fluxes}
  \begin{split}
    \mc F_A \equiv \lim_{\Delta w \to \infty} \frac{\<z(\tau')\>_{\tau'}}{16\pi C_g \Delta w} \bigg[&\int_{\scri^+ (\Delta w)} \bs \omega\{\bs h, \nabla_A \bs h\} \\
    &+ \int_{\scrh^+ (\Delta w)} \bs \omega\{\bs h, \nabla_A \bs h\}\bigg],
  \end{split}
\end{equation}
where $\scri^+ (\Delta u)$ and $\scrh^+ (\Delta v)$ are portions of future null infinity and the future horizon of length $\Delta u$ and length $\Delta v$, respectively.
Note that this conserved current is defined in terms of metric perturbations, and so to obtain a more practical flux-balance law one should rewrite the symplectic current in terms of Teukolsky variables by metric reconstruction, as was done for example in~\cite{Grant2020}.
A trick for circumventing such a calculation, using the results of~\cite{Isoyama2018}, is laid out in~\cite{Mathews2025}.

In the next two sections, we consider the implications of Eq.~\eqref{eqn:result} in the case of non-spinning particles, and for spinning particles to linear order in spin.

\subsection{Non-spinning particles} \label{sec:geodesic_kerr}

We first consider the non-spinning case.
Here, the 8-dimensional phase space is integrable: for the background motion, there exists a set of four constants of motion $P_\alpha$ which are linearly independent, so that
\begin{equation}
  (\ud P_1) \wedge \cdots \wedge (\ud P_4) \neq 0,
\end{equation}
and are in involution, so that
\begin{equation}
  \{P_\alpha, P_\beta\} = 0.
\end{equation}
Two of these constants of motion, $E$ and $L_z$, are linear in $p_a$ and related to the isometries of the Kerr spacetime:
\begin{equation}
  E \equiv -t^a p_a, \qquad L_z \equiv \varphi^a p_a,
\end{equation}
where $t^a$ and $\varphi^a$ are the Killing vectors generating $t$ and $\varphi$ translations, respectively, in Boyer-Lindquist coordinates.
The other two constants, $m^2$ and the Carter constant $K$, are quadratic, and are given by
\begin{equation}
  m^2 \equiv -g^{ab} p_a p_b, \qquad K \equiv K_{ab} p^a p^b,
\end{equation}
where $K_{ab}$ is a second-rank Killing tensor satisfying
\begin{equation}
  \nabla_{(a} K_{bc)} = 0
\end{equation}
(note that the metric $g_{ab}$ is also, trivially, a second-rank Killing tensor).

Since this system is integrable, there exist~\cite{Arnold1989, Fiorani2003, Hinderer2008} action-angle variables $(\vartheta^\alpha, J_\alpha)$, which are canonical in the sense that the only nonzero Poisson bracket is determined by
\begin{equation}
  \{\vartheta^\alpha, J_\beta\} = \delta^\alpha{}_\beta,
\end{equation}
and $H$ is only a function of $J_\alpha$, so that Hamilton's equations become
\begin{equation}
  \dot \vartheta^\alpha = \frac{\partial H}{\partial J_\alpha}, \qquad \dot J_\alpha = 0,
\end{equation}
where $\partial H/\partial J_\alpha$ are the frequencies associated with the action variables.
Generically, one can instead use the variables $(\vartheta^\alpha, P_\alpha)$, which have similar properties, except they fail to be exactly canonical:
\begin{equation} \label{eqn:quasicanonical_geodesic}
  \{\vartheta^\alpha, P_\beta\} = \frac{\partial P_\beta}{\partial J_\alpha} \equiv A^\alpha{}_\beta,
\end{equation}
This set of coordinates is sufficient for the discussion in this paper.

In the coordinates $(\vartheta^\alpha, P_\alpha)$, the fact that
\begin{equation}
  \dot \vartheta^\alpha = \nu^\alpha (P), \qquad \dot P_\alpha = 0,
\end{equation}
for some frequencies $\nu^\alpha$, implies that
\begin{subequations}
  \begin{align}
    P_\alpha (\tau') &= P_\alpha (\tau), \\
    \vartheta^\alpha (\tau') &= \vartheta^\alpha (\tau) + (\tau' - \tau) \nu^\alpha (P).
  \end{align}
\end{subequations}
Using Eq.~\eqref{eqn:coordinates} for the Hamilton propagator in coordinates, it follows that
\begin{equation} \label{eqn:geodesic_propagator}
  \Upsilon^\aleph{}_\beth (\tau', \tau) = \begin{pmatrix}
    \delta^\alpha{}_\beta & (\tau' - \tau) \frac{\partial \nu^\alpha}{\partial P_\beta} \\
    0 & \delta^\beta{}_\alpha
  \end{pmatrix}.
\end{equation}
As such, by taking the $P_\alpha$ component of Eq.~\eqref{eqn:result}, one finds from Eqs.~\eqref{eqn:quasicanonical_geodesic} and~\eqref{eqn:geodesic_propagator} that
\begin{equation} \label{eqn:almost_true_result}
  \begin{split}
    \left\<\var \dot P_\alpha (\tau')\right\>_{\tau'} = -A^\beta{}_\alpha \bigg\{&(\partial_{\vartheta^\beta})^A \mc F_A \\
    &+ \frac{\partial \<\mc H[\Upsilon_{\tau' - \tau} (X)]\>_{\tau'}}{\partial \vartheta^\beta}\bigg\}.
  \end{split}
\end{equation}
In order to derive Eq.~\eqref{eqn:true_result}, we then need only show that the second term on the right-hand side of this equation vanishes.
This follows from an argument analogous to that in Secs.~2.3 and~3.2 of~\cite{Isoyama2018}, at least in the case of non-resonant background orbits.
There is another, more straightforward way of understanding this argument which we present below.

First, note that we can Fourier expand $\var H(X, \bar X)$ in these coordinates as\footnote{That this is a Fourier series (and not a Fourier transform) for the non-compact angle variable follows from the fact that the background orbit from which this quantity is constructed is (multi)periodic.
  In particular, we need to assume that the background orbit is \emph{bound}.}
\begin{equation} \label{eqn:bifourier}
  \var H(X, \bar X) = \sum_{\bs n, \bar{\bs n}} H_{\bs n, \bar{\bs n}} (P, \bar P) e^{i(n_\alpha \vartheta^\alpha + \bar n_\alpha \bar \vartheta^\alpha)}.
\end{equation}
However, note that the dependence of $\var H$ on $\bar X$ is only through the full phase-space trajectory which passes through $\bar X$, and so
\begin{equation}
  \begin{split}
    \var H(X, \bar X) &= \var H[X, \Upsilon_\Delta (\bar X)] \\
    &= \sum_{\bs n, \bar{\bs n}} H_{\bs n, \bar{\bs n}} (P, \bar P) e^{i\{n_\alpha \vartheta^\alpha + \bar n_\alpha [\bar \vartheta^\alpha + \Delta \bar \nu^\alpha (\bar P)]\}},
  \end{split}
\end{equation}
for any $\Delta$.
Comparing this with Eq.~\eqref{eqn:bifourier}, using the linear independence of the complex exponential, and using the non-resonance condition
\begin{equation}
  n_\alpha \neq 0\; \implies\; n_\alpha \nu^\alpha (P) \neq 0,
\end{equation}
we therefore find that only $H_{\bs n, 0}$ can be non-zero.
As such, we can write
\begin{equation}
  \mc H[\Upsilon_{\tau' - \tau} (X)] = \sum_{\bs n} \mc H_{\bs n} (P) e^{i n_\alpha [\vartheta^\alpha + (\tau' - \tau) \nu^\alpha (P)]};
\end{equation}
in general, if the only nonzero terms in the Fourier expansion \emph{weren't} those with $\bar n^\alpha = 0$, there could be \emph{different} quadruples of integers appearing in front of the $\vartheta^\alpha$ and $\nu^\alpha (P)$ terms.
Upon averaging this equation over $\tau'$, and once again using the non-resonance condition, we find that
\begin{equation}
  \<\mc H[\Upsilon_{\tau' - \tau} (X)]\>_{\tau'} = \mc H_0 (P),
\end{equation}
from which it follows that the second term in Eq.~\eqref{eqn:almost_true_result} vanishes, and so Eq.~\eqref{eqn:true_result} holds.

We conclude this section by considering the \emph{angle} components of Eq.~\eqref{eqn:result}.
Applying this equation na\"ively, and using Eqs.~\eqref{eqn:quasicanonical_geodesic} and~\eqref{eqn:geodesic_propagator}, we would find that
\begin{equation} \label{eqn:conjugate_result}
  \begin{split}
    \left\<\frac{\ud}{\ud \tau'}\right. &\left.\left[\var \vartheta^\alpha (\tau') - (\tau' - \tau) \frac{\partial \nu^\alpha}{\partial P_\beta} \var P_\beta (\tau')\right]\right\>_{\tau'} \\
    &= A^\alpha{}_\beta \left\{(\partial_{P_\beta})^A \mc F_A + \frac{\partial}{\partial P_\beta} \<\mc H[\Upsilon_{\tau' - \tau} (X)]\>\right\}.
  \end{split}
\end{equation}
This equation, however, is not entirely correct.
First, the left-hand side diverges: the quantity inside the average does not have a well-defined average, as one can show that it grows linearly with $\tau' - \tau$.
This is also reflected in the right-hand side as well: the flux term \emph{also} diverges, as $\partial_{P^\alpha} h_{a'b'} (X)$ grows linearly with $\tau' - \tau$, and so the assumption that $\nabla_A h_{a'b'} (X)$ stays finite no longer holds.

There are two possible ways around this issue.
First, note that, in general, there can exist points in phase space where $A^\alpha{}_\gamma \partial \nu^\beta/\partial P_\gamma$ is no longer invertible: this will occur where the mapping between action variables and frequencies is no longer injective; that is, along \emph{isofrequency} curves.
As such, there will exist at least one dual vector $V_\alpha$ such that
\begin{equation}
  V_\alpha A^\alpha{}_\beta \frac{\partial \nu^\gamma}{\partial P_\beta} = 0,
\end{equation}
and so contracting $V_\alpha$ into both sides of Eq.~\eqref{eqn:conjugate_result} would give an expression which is well-defined, although it would not be a true flux-balance law, as it would still involve the local self-force through the conservative Hamiltonian.
As another option, it may be possible to recover useful results by generally considering both sides of Eq.~\eqref{eqn:conjugate_result} in a power series in $\tau' - \tau$, and equating terms of equal powers.
As this begins to stretch the notion of what one can mean by a ``flux-balance law'', we leave a more careful exploration of this possibility to future work.

\subsection{Spinning particles} \label{sec:spinning_kerr}

We now consider the linear-in-spin case.
There has been much work on studying spinning test bodies in Kerr in terms of conserved quantities and action-angle variables in the Hamiltonian formulation~\cite{Witzany2019, Ramond2022b, Compere2023a, Ramond2024}; the description in this section is primarily based on that in~\cite{Ramond2022b}.

Here, there do \emph{not} exist a known set of action-angle variables on the 14-dimensional phase space, for the background motion.
This is due to the fact that, considering the full 14-dimensional phase space, the motion is constrained to lie on a 10-dimensional physical phase space $\mc P$.
While there exist action-angle variables on $\mc P$ itself, when moving off of $\mc P$, some conserved quantities are no longer conserved.
This significantly complicates the analysis, in particular the computation of the Hamilton propagator.

Explicitly, let us denote the Poisson bivector on $\mc P$ by $\ul \Pi^{AB}$, and the corresponding bracket by $\ul{\{\cdot, \cdot\}}$.
The results of~\cite{Ramond2022b} imply that there exist a set of action and angle variables $(\Theta^\clubsuit, \mc J_\clubsuit)$ on $\mc P$ that have the following Poisson bracket structure:
\begin{equation}
  \ul{\{\Theta^\clubsuit, \mc J_\spadesuit\}} = \delta^\clubsuit{}_\spadesuit,
\end{equation}
where we are using card suits ($\clubsuit$, $\spadesuit$, etc.) for five-dimensional indices on half of the ten-dimensional phase space $\mc P$.

Now, the action variables $\mc J_\clubsuit$ are constructed from five constants of motion: the first four are $m^2$, $E$, $L_z$, and $K$, but they are modified relative to their original definitions in the absence of spin.
The modification to the energy and angular momentum is well-known~\cite{Rudiger1981, Harte2014}, and given by
\begin{equation}
  E = -\left(p_a t^a + \frac{1}{2} S^{ab} \nabla_a t_b\right), \quad L_z = p_a \varphi^a + \frac{1}{2} S^{ab} \nabla_a \varphi_b.
\end{equation}
The modification of $K$ is less trivial, and related to the existence of a Killing-Yano tensor $Y_{ab}$ in the Kerr spacetime from which the Killing tensor $K_{ab}$ in the Kerr spacetime can be constructed~\cite{Walker1970}:
\begin{equation}
  K_{ab} = Y_{ac} Y^c{}_b.
\end{equation}
The modification to $K$ is then given by~\cite{Rudiger1983, Ramond2022b}
\begin{equation}
  K = p_a p_b K^{ab} - 12  S^{d[b} p^{c]} Y^e{}_{c} \nabla_{[d} Y_{eb]}.
\end{equation}
The final constant of motion, the R\"udiger constant $Y$, is given by~\cite{Rudiger1981, Ramond2022b}
\begin{equation}
  Y = \frac{1}{4} \epsilon_{abcd} Y^{ab} S^{cd}.
\end{equation}

In addition to these sets of variables, we also consider a set of four ``constraint'' variables $C^\alpha$.
Three of these are just the components of $d^a$, which we set to zero on $\mc P$.
The fourth is a Casimir invariant $C_\circ$, which we define by~\cite{Ramond2022b}
\begin{equation}
  C_\circ \equiv \frac{1}{2} \eta_{\hat \alpha \hat \gamma} \eta_{\hat \beta \hat \delta} \hat S^{\hat \alpha \hat \beta} \hat S^{\hat \gamma \hat \delta} - S_\circ,
\end{equation}
for any constant $S_\circ$ set by the initial value of the first term such that $C_\circ = 0$ for the physical background motion.
It can be shown that this quantity obeys
\begin{equation}
  \Pi^{AB} (\ud C_\circ)_A = 0,
\end{equation}
which shows that $\Pi^{AB}$ is degenerate; as remarked above in Sec.~\ref{sec:formalism}, this poses no issue.
Similarly, there is another Casimir invariant $C_\star$ defined by~\cite{Ramond2022b}
\begin{equation}
  C_\star \equiv \frac{1}{8} \hat \epsilon_{\hat \alpha \hat \beta \hat \gamma \hat \delta} \hat S^{\hat \alpha \hat \beta} S^{\hat \gamma \hat \delta},
\end{equation}
where $\hat \epsilon_{\hat \alpha \hat \beta \hat \gamma \hat \delta}$ is the Levi-Civita symbol.
Note that $C_\star$ vanishes on the physical background phase space as well, and in fact by a suitable choice of coordinates $C_\star$ can be chosen to be one of the components of $d^a$.
As such, explicitly, the components of $C^\mu$ are given by $C_\circ$, $C_\star$, and the two remaining independent components of $d^a$.

The key property of the R\"udiger constant is that it is the \emph{only} one of these constants in the Kerr spacetime which requires that the spin supplementary condition hold~\cite{Ramond2022b, Compere2023a}: denoting the collection of the first four constants of motion by $P_\alpha$, as above, we find that
\begin{equation}
  \dot P_\alpha = 0, \qquad \dot Y = O(\bs C).
\end{equation}
Similarly, while
\begin{equation}
  \{P_\alpha, C^\beta\} = O(\bs C),
\end{equation}
we have that
\begin{equation}
  \{Y, P_\alpha\} \neq 0, \qquad \{Y, C^\alpha\} \neq 0,
\end{equation}
even on $\mc P$.
However, there exists an alternative R\"udiger constant $\tilde Y$, defined by\footnote{The author thanks Paul Ramond for pointing out that such a quantity exists.}
\begin{equation}
  \tilde Y = \frac{1}{2m^2} \varepsilon_{abcd} p^b S^{cd} p_e Y^{ae},
\end{equation}
such that $Y$ and $\tilde Y$ agree on $\mc P$, but that
\begin{equation}
  \{\tilde Y, P_\alpha\} = O(\bs C), \qquad \{\tilde Y, C^\alpha\} = O(\bs C).
\end{equation}
Similarly, we still have that
\begin{equation}
  \dot{\tilde Y} = O(\bs C).
\end{equation}

In summary, what we now have is the following situation: we can work in a set of coordinates near $\mc P$ given by
\begin{equation}
  X^\aleph = \begin{pmatrix}
    \vartheta^\alpha \\
    \upsilon \\
    P_\alpha \\
    \tilde Y \\
    C^\alpha
  \end{pmatrix},
\end{equation}
where we have split the angle variables $\Theta^\clubsuit$ into $\vartheta^\alpha$ and $\upsilon$, such that
\begin{itemize}

\item the only nonzero components of $\Pi^{AB} (\ud P_\alpha)_B$ and $\Pi^{AB} (\ud \tilde Y)_B$, once restricted to $\mc P$, are given by
  \begin{subequations} \label{eqn:P_poisson}
    \begin{align}
      \{\vartheta^\alpha, P_\beta\} &= A^\alpha{}_\beta, &\{\upsilon, P_\alpha\} &= A_\alpha, \\
      \{\vartheta^\alpha, \tilde Y\} &= A^\alpha, &\{\upsilon, Y\} &= A;
    \end{align}
  \end{subequations}

\item the only nonzero components of $\Pi^{AB} (\ud C^\alpha)_B$, once restricted to $\mc  P$, are given by
  \begin{subequations} \label{eqn:C_poisson}
    \begin{align}
      \{C^\alpha, C^\beta\} &= B^{\alpha\beta}, \\
      \{\vartheta^\alpha, C^\beta\} = D^{\alpha\beta}, \quad &\quad \{\upsilon, C^\alpha\} = D^\alpha;
    \end{align}
  \end{subequations}
  and

\item off of $\mc P$, Hamilton's equations imply that
  \begin{subequations}
    \begin{align}
      \dot P_\alpha &= 0, \label{eqn:P_dot} \\
      \dot{\tilde Y} &= E_\alpha C^\alpha, \label{eqn:rudiger_dot} \\
      \dot C^\alpha &= F^\alpha{}_\beta C^\beta, \label{eqn:C_dot}
    \end{align}
  \end{subequations}
  where $E_\alpha$ and $F^\alpha{}_\beta$ are functions of all of the variables on $\mc P$.

\end{itemize}

In the above paragraph, we have made no assumption about the splitting of the five angle variables $\Theta^\clubsuit$ into $\vartheta^a$ and $\upsilon$.
One final assumption which we make is that this angle variable is such that the retarded metric perturbation $h_{ab}$, in terms of its dependence on initial angle variables, can be split into two pieces,
\begin{equation}
  h_{ab} = \sum_{\bs n} h^{\bs n}_{ab} e^{in_\alpha \vartheta^\alpha} + \sum_{\bs n, N} h^{\bs n, N}_{ab} e^{i(n_\alpha \vartheta^\alpha + N \upsilon)},
\end{equation}
where $h^{\bs n, N}_{ab}$ is $O(S)$.
This is the case, for example, in the multiscale expansion in~\cite{Mathews2025}.
As such, we have that
\begin{equation}
  (\partial_\upsilon)^A \mc F_A = O(S^2),
\end{equation}
and so terms proportional to $(\partial_\upsilon)^A \mc F_A$ can be neglected.

We now discuss the consequences of the existence of these coordinates.
First, from Eqs.~\eqref{eqn:P_poisson} and~\eqref{eqn:P_dot}, it follows that Eq.~\eqref{eqn:true_result} still holds:
\begin{equation} \label{eqn:miraculous_result}
  \left\<\dot P_\alpha(\tau')\right\>_{\tau'} = -A^\beta{}_\alpha (\partial_{\vartheta^\beta})^A \mc F_A.
\end{equation}
This implies that one can still write down flux-balance laws for the $P_\alpha$, which in this case are given by $m^2$, $E$, $L_z$, and the generalization of the Carter constant $K$.
While such flux-balance laws have been known for $E$ and $L_z$~\cite{Akcay2019, Mathews2021}, this shows that straightforward generalizations exist for two more of the constants of motion.
Note that, despite appearing to have the same form as Eq.~\eqref{eqn:true_result}, the matrix $A^\alpha{}_\beta$ which appears in Eq.~\eqref{eqn:miraculous_result} can depend on spin, and so the explicit expressions will not be exactly the same.
Splitting the action variables $\mc J_\clubsuit$ into action variables $J_\alpha$ and $\Upsilon$, we have that
\begin{equation}
  A^\alpha{}_\beta = \frac{\partial P_\beta}{\partial J_\alpha},
\end{equation}
as before in Eq.~\eqref{eqn:quasicanonical_geodesic}.
While an explicit full set of action variables has recently been defined in~\cite{Witzany2024}, we leave explicit computations of $A^\alpha{}_\beta$, and therefore an explicit form of the flux-balance law, to future work~\cite{Mathews2025}.

That we were able to derive Eq.~\eqref{eqn:miraculous_result} is due to the ``miraculous'' coincidence that both $m^2$ and $K$, like $E$ and $L_z$ but \emph{unlike} $\tilde Y$, are still conserved even when the spin-supplementary condition does not hold.
Determining whether there exists a flux-balance law for $\tilde Y$ is far more difficult.
In fact, since generically the constraints \emph{also} evolve, determining the average evolution of $\tilde Y$ is not even sufficient for determining the total, average motion.
To see why the constraints evolve, note that while $\breve C^\alpha (\varepsilon)$ (that is, the constraints defined using the effective metric) vanishes when evaluated on the exact phase space trajectory $\Gamma(\varepsilon)$, and $C^\alpha$ (the constraints in the background metric) vanishes when evaluated on the background trajectory $\Gamma$, the quantity $\var C^\alpha$ is defined by
\begin{equation}
  \var C^\alpha (\tau) \equiv \left.\var \Gamma^A \nabla_A C^\alpha\right|_{\Gamma(\tau)} = \left.\frac{\partial C^\alpha [\Gamma(\tau; \varepsilon)]}{\partial \varepsilon}\right|_{\varepsilon = 0}, \vspace{\belowdisplayskip}
\end{equation}
which does \emph{not} vanish; it would only vanish if $C^\alpha$ on the right-hand side were replaced with $\breve C^\alpha$.

Returning to the flux-balance law for the R\"udiger constant, the first, and greatest complication, is that the Hamilton propagator no longer has the simple structure that it has in Eq.~\eqref{eqn:geodesic_propagator}, since solving for the Hamilton propagator requires solving the equations of motion to linear order, which will necessarily introduce constraint terms.
Explicitly, Eq.~\eqref{eqn:rudiger_dot} implies that
\begin{equation}
  \tilde Y(\tau') = \tilde Y(\tau) + \int_\tau^{\tau'} E_\alpha (\tau'') C^\alpha (\tau'') \ud \tau'' + O(\bs C^2).
\end{equation}
Similarly, we find from Eq.~\eqref{eqn:C_dot} that
\begin{equation} \label{eqn:C_evolve}
  C^\alpha (\tau') = U^\alpha{}_\beta (\tau', \tau) C^\beta (\tau) + O(\bs C^2),
\end{equation}
where $U^\alpha{}_\beta (\tau', \tau)$ is the solution to
\begin{equation}
  \frac{\partial}{\partial \tau'} U^\alpha{}_\beta (\tau', \tau) = F^\alpha{}_\gamma (\tau') U^\gamma{}_\beta (\tau', \tau)
\end{equation}
such that
\begin{equation}
  U^\alpha{}_\beta (\tau, \tau) = \delta^\alpha{}_\beta.
\end{equation}
As such, writing
\begin{equation}
  V_\alpha (\tau', \tau) \equiv \int_\tau^{\tau'} E_\beta (\tau'') U^\beta{}_\alpha (\tau'', \tau) \ud \tau'',
\end{equation}
we find
\begin{equation} \label{eqn:Y_evolve}
  \tilde Y(\tau') = \tilde Y(\tau) + V_\alpha (\tau', \tau) C^\alpha (\tau) + O(\bs C^2).
\end{equation}

Due to the complications appearing in Eqs.~\eqref{eqn:Y_evolve} and~\eqref{eqn:C_evolve}, the $\tilde Y$ or $C^\alpha$ components of the left-hand side of Eq.~\eqref{eqn:result} are no longer simple.
Using the Poisson bracket structure in Eqs.~\eqref{eqn:P_poisson} and~\eqref{eqn:C_poisson}, we find that
\begin{widetext}
\begin{subequations} \label{eqn:confusing_result}
  \begin{align}
    \left\<\frac{\ud}{\ud \tau'} \left[\var \tilde Y(\tau') + V_\alpha (\tau, \tau') \var C^\alpha (\tau')\right]\right\>_{\tau'} &= -A^\alpha (\partial_{\vartheta^\alpha})^A \mc F_A, \label{eqn:rudiger_result} \\
    \left\<\frac{\ud}{\ud \tau'} \left[U^\alpha{}_\beta (\tau, \tau') \var C^\beta (\tau')\right]\right\>_{\tau'} &= [B^{\alpha\beta} (\partial_{C^\beta})^A - D^{\alpha\beta} (\partial_{\vartheta^\beta})^A] \mc F_A + B^{\alpha\beta} \frac{\partial}{\partial C^\beta} \<\mc H(X, \tau')\>_{\tau'}, \label{eqn:constraint_result}
  \end{align}
\end{subequations}
\end{widetext}
where [much like for Eq.~\eqref{eqn:true_result}] we are assuming that we are off resonance.

This equation shows two important features: first, that the question ``what is the average rate of change of $\tilde Y$ and $C^\alpha$'' is \emph{not} covariant on phase space: one should instead ask questions about the left-hand sides of these equations.
It should, however, be noted that it is not clear if these quantities even remain finite, particularly in the case of Eq.~\eqref{eqn:constraint_result} [Eq.~\eqref{eqn:rudiger_result} is saved by the fact that the right-hand side is clearly finite].
We will leave to future work the question of whether the quantities on the left-hand side of these equations are useful for practical calculations.
The second important feature is that not all terms on the right-hand side are purely asymptotic: they depend on the average value of the conservative Hamiltonian.
If this remains true with a more careful exploration of this system, then this equation is not a ``true'' flux-balance law.

\section{Discussion} \label{sec:discussion}

In this paper, under a very broad set of assumptions (not even restricting to the Kerr spacetime), we have shown that a type of quasi-local flux-balance law exists for motion under the gravitational self-force, both for non-spinning and for spinning particles.
This flux-balance law is, unlike many derived previously in the literature, explicitly constructed from a conserved current in the background spacetime.
By restricting to a Kerr background, we have moreover confirmed the existence of true flux-balance laws, only involving asymptotic metric perturbations, for non-resonant orbits and for non-spinning bodies~\cite{Isoyama2018}.
Moreover, we have shown that there are difficulties in extending this flux-balance law to linear order in spin: while there exist flux-balance laws for the generalizations of the conserved quantities for geodesic motion ($m^2$, $E$, $L_z$, and the Carter constant $K$), the R\"udiger constant $Y$ and the constraint (that is, spin-supplementary-condition) violation terms do not seem to have flux-balance laws.

Our result here, however, should not be considered as a no-go theorem for true flux-balance laws for spinning bodies in the Kerr spacetime.
We present here some options for potentially recovering flux-balance laws, to be explored in future work.
First, our results, at this stage, are not entirely explicit: in this paper, the results have mostly depended on the structure of many of the equations, and not in the particular expressions which appear, for example in the various quantities appearing on the right-hand sides of Eqs.~\eqref{eqn:C_poisson}, \eqref{eqn:rudiger_dot}, and~\eqref{eqn:C_dot}.
It is possible that, once an explicit calculation is completed, many of the offending terms in Eq.~\eqref{eqn:confusing_result} will vanish, or simply be higher order in spin.
This is suggested by recent results showing that $\tilde Y$, or rather a quantity trivially related to it (the ``parallel component of the spin vector''), does not evolve on average in the linear-in-spin approximation~\cite{Skoupy2024, Mathews2025}.
That is, while the results of this paper imply that there may not be a meaningful sense in which $\tilde Y$ possesses a flux-balance law, it may not matter.

While this could potentially resolve issues with the evolution of $\tilde Y$, one remaining issue would be in determining the evolution of the constraint violation terms $C^\alpha$.
Note that these terms arise due to the fact that we are using the Tulczyjew-Dixon spin supplementary condition associated with the effective metric, but measuring the deviations from this condition as computed from the \emph{background} metric.
Using this choice of spin supplementary condition is the most natural, from the perspective of the generalized equivalence principle.
Moreover, it is a necessary ingredient in deriving the Hamiltonian alignment condition and therefore our main result in Eq.~\eqref{eqn:result}.
However, it may be easier to use the \emph{background} Tulczyjew-Dixon spin supplementary condition, and simply amend our result accordingly.
In this case, the constraint violation terms should not appear.
Similarly, it may be useful to consider the Hamiltonian formulations for alternative spin supplementary conditions considered in~\cite{Witzany2018}.
Since the choice of spin-supplementary condition corresponds only to an arbitrary choice of center-of-mass worldline for an extended body (in particular, there exist formulations which do not require an explicit choice~\cite{Harte2014}), it is plausible that the asymptotic metric perturbation may be independent of $C^\alpha$, and so its averaged evolution is not required.

Even if true flux-balance laws do not exist for the R\"udiger constant $Y$ and the constraint violation terms $C^\alpha$, flux-balance laws for the remaining constants of motion will still be useful.
Moreover, expressions such as Eq.~\eqref{eqn:confusing_result} will still hold, even if they cannot be used as a computational shortcut.
At the very least, they will provide quasi-local checks of the validity of more direct calculations of the gravitational self-force, and so may still be of some use.

Finally, a future direction for this work is to extend the results to second order in $\varepsilon$.
Here, there is a major conceptual issue with the calculation in this paper: the first order self-force as discussed in this paper is first-order in the sense of a Taylor expansion in powers of $\varepsilon$.
However, it is known~\cite{Pound2009} that such an expansion breaks down on long times: even if the exact trajectory $\Gamma(\varepsilon)$ is initially close to the background worldline $\Gamma$, it will in general diverge on long timescales.
This is the motivation for using a \emph{self-consistent} approach to gravitational self-force, where one uses the exact trajectory $\Gamma(\varepsilon)$ as the source for the metric perturbation, and self-consistently solves for both order-by-order.
The key difference between this approach and usual perturbative expansions is that the exact trajectory is \emph{never} expanded, and so the coefficients at each order in $\varepsilon$ are no longer $\varepsilon$-independent.

This approach is, in a certain sense, the most ``elegant'', and in fact, preliminary investigations carried out while working on this paper suggest that many of the core features required for flux-balance laws, such as the Hamiltonian alignment condition, already hold due to the existence of the Detweiler stress-energy tensor (see~\cite{Upton2021}) at second order.
However, there is a deep conceptual issue that means that the self-consistent approach seems unlikely to be useful for flux-balance laws: since the self-consistent approach uses an exact trajectory $\Gamma(\varepsilon)$, it follows that $\gamma(\varepsilon)$ will necessarily plunge into the black hole.
This means that one can no longer perform infinite proper-time averages over bound motion.
The issue here is essentially that flux-balance laws are very nonlocal in time (in addition to space!), and so the self-consistent approach is unlikely to yield useful results.

On the other hand, there is a different approach to the gravitational self-force, known as a multiscale expansion~\cite{Kevorkian2012, Hinderer2008, Pound2021}.
The general idea is that, by introducing additional (``slow time'') variables, one can parametrize the exact trajectory in such a way that the trajectory, at fixed slow time, can be expanded in a usual Taylor expansion in terms of $\varepsilon$-independent coefficients, and the behavior on long timescales is captured by the evolution of the slow time variables themselves.
This gives the best features of both usual perturbative expansions and the self-consistent approach, but as we will show in upcoming work~\cite{isometries20XX}, it comes with a catch when one attempts to formulate flux-balance laws.
With the introduction of the slow time variables, the perturbative field equations at each order contain derivatives with respect to these variables \{see the discussion in Sec.~7.1.1 of~\cite{Pound2021}, in particular the second line of Eq.~(396)\}.
As such, there is no longer a sense in which the perturbative Einstein equations hold at second order in $\varepsilon$.
By breaking the perturbative Einstein equations, the machinery for constructing conserved currents in this paper will necessarily fail.
While that can be amended through careful choices of additional, correcting currents~\cite{isometries20XX}, the ultimate conclusion is that one will still need to compute a portion of the second-order metric perturbation on the worldline.
While this may diminish the utility of flux-balance laws at second order, it remains to be seen how fatal this issue may ultimately be.

\section{Acknowledgments}

The author thanks Soichiro Isoyama, Adam Pound, and Paul Ramond for many insightful discussions, Sam Upton for clarifying the status of the self-force formalism in the presence of spin, and Josh Mathews for sharing a preliminary version of~\cite{Mathews2025}.
The author also thanks Francisco Blanco, Soichiro Isoyama, Josh Mathews, Adam Pound, Paul Ramond, and Vojt\v ech Witzany for valuable feedback on early drafts.
This work was supported by the Royal Society under grant number {RF\textbackslash ERE\textbackslash 221005}.

\bibliographystyle{utphys}
\bibliography{refs}

\end{document}